\documentclass[5p, authoryear,10pt]{elsarticle}
\usepackage{float}
\usepackage[fleqn]{amsmath}
\usepackage{amssymb,amsfonts,amsmath}
\usepackage{setspace}
\newcommand{\vect}[1]{\textbf{#1}}
\newcommand{\tens}[1]{\boldsymbol{#1}}
\newcommand{\Div}{\pmb{\nabla} \cdot}
\newcommand{\Grad}{\pmb{\nabla}}
\usepackage[bookmarksnumbered,colorlinks,linkcolor=blue]{hyperref}	
\usepackage{soul}
\usepackage{xcolor}
\usepackage{enumitem}
\usepackage[normalem]{ulem}

\usepackage{lineno}

\definecolor{darkblue}{RGB}{0,84,147}
\definecolor{darkgreen}{RGB}{0,147,84}

\journal{Icarus}

\begin{document}

\begin{frontmatter}

\title{Time scales of small body differentiation}

\author[a]{Marc Monnereau}
\author[a,b]{Jérémy Guignard} 
\author[a,c]{Adrien Néri} 
\author[a]{Michael J. Toplis} 
\author[a]{Ghylaine Quitté} 
\address[a]{IRAP, University of Toulouse, CNRS, Toulouse, France}
\address[b]{Now at ICMCB, CNRS, Université de Bordeaux, Bordeaux, France}
\address[c]{Now at BGI, University of Bayreuth, Bayreuth, Germany}

\begin{abstract}

The petrologic and geochemical diversity of meteorites is a function of the bulk composition of their parent bodies, but also the result of how and when internal differentiation took place.
Here we focus on this second aspect considering the two principal parameters involved: size and accretion time of the body. 
We discuss the interplay of the various time scales related to heating, cooling and drainage of silicate liquids. 
Based on two phase flow modelling in 1-D spherical geometry, we show that drainage time is proportional to two independent parameters:  $\mu_m/R^2$, the ratio of the matrix viscosity to the square of the body radius and $\mu_f/a^2$, the ratio of the liquid viscosity to the square of the matrix grain size. 
We review the dependence of these properties on temperature, thermal history and degree of melting, demonstrating that they vary by several orders of magnitude during thermal evolution.
These variations call into question the results of two phase flow modelling of small body differentiation that assume constant properties.
For example, the idea that liquid migration was efficient enough to remove $^{26}$Al heat sources from the interior of bodies and dampen their melting  \cite[e.g.][]{Moskovitz2011, Neumann2012} relies on percolation rates of silicate liquids overestimated by six to eight orders of magnitude.
In bodies accreted during the first few million years of solar-system history, we conclude that drainage cannot prevent the occurrence of a global magma ocean. These conditions seem ideal to explain the generation of the parent-bodies of iron meteorites. A map of the different evolutionary scenarios of small bodies as a function of size and accretion time is proposed.

\end{abstract}



\begin{keyword}
Asteroid \sep differentiation \sep thermal history \sep melt migration
\end{keyword}
\end{frontmatter}


\newpage
\section{Introduction}

The meteoritic record constitutes a well-preserved selection of samples that allow us to shed light on the differentiation processes that occurred on small rocky bodies that accreted early in solar-system history. From the suite of specimens available, there is an obvious link between the overall degree of melting of the parent bodies and the degree of differentiation. 
Unlike larger planets, the main heat source of these small bodies is the energy produced by the decay of short-lived radionuclides such as $^{26}$Al: the earlier a body accreted, the greater the heating potential and melting degree. 
The parent bodies of the iron meteorites accreted very early \citep[$< 0.3$\,Myr after CAIs ---CAI stands for Calcium Aluminum Inclusions, the most refractory objects whose condensation defines the age of the solar system---][]{kruijer2014}, with the potential to produce large-scale magma oceans that ensured an efficient metal-silicate differentiation through an iron-rain scenario \cite[e.g.][]{Rubie2003}.  
Conversely, chondritic parent bodies accreted much later \citep[$>$ 2 Myr after CAIs][]{sugiura2014} such that peak temperatures were not high enough to reach the iron-sulfur or silicate solidi, preventing any differentiation. 
In between these two differentiation endmembers are the primitive achondrites that accreted early enough $\approx$1.3 Myr after CAIs to be able to cross the silicate solidus but late enough so that their degrees of partial melting did not exceed 20 vol\% \citep{sugiura2014}. Under these conditions, these samples experienced partial differentiation, with the loss of basaltic and sulfur-rich metallic melts.

At first sight, a difference in accretion dates thus provides a satisfactory explanation for the different degrees of differentiation found in the meteoritic record. 
However, body size will play an important role too. Indeed, for the same accretion time, a smaller body will dissipate its heat with greater efficiency than a larger one, thus yielding a different heating potential and a lower peak temperature for the former.
Attempts to take this effect into account are scarce and restrained to a handful of samples for which precise anchor-points in the cooling history have been determined (e.g. \citeauthor{Breton2015}, \citeyear{Breton2015} for Tafassasset and \citeauthor{Neumann2018}, \citeyear{Neumann2018} for the acapulcoite-lodranite parent body).
The present paper aims at describing the general pattern of evolution and differentiation that allows the fate of a parent body to be predicted as a function of its size and accretion time.

 An early attempt in this direction was made by \cite{Moskovitz2011} who used simple considerations of the time scales involved to study the consequences of silicate melt migration on the thermal evolution of small bodies.
In particular, they popularised the idea that low viscosity melt ($<1$\,Pas) can transport $^{26}$Al heat sources to the surface on time scales shorter than their mean half-live of radioactive decay. 
They stressed the role of melt viscosity in controlling the degree of melting, mentioning that, above 1\,Pas, melt content in bodies accreted before 1.5\,Myr should exceed the 50\% threshold capable of triggering the generation of a magma ocean --- 
above about 50\% liquid, the rigid silicate framework dissociates and the matrix becomes a dense suspension of crystals in a liquid; the viscosity of the solid-liquid mixture changes from that of a solid to that of a liquid \citep[e.g.][]{Solomatov2015}.
Using the more complex approach offered by  two-phase flow numerical modelling, \cite{Lichtenberg2019} found a complementary result. They emphasised the role of grain size, indicating inefficient liquid migration with a matrix grain size below one millimeter (conclusion drawn making the assumption of a melt viscosity of the order of 1 Pas). 
In fact, the percolation rate is partially controlled by the ratio $\mu_f/a^2$ where $\mu_f$ is the melt viscosity and $a$ is the grain size, such that efficient drainage is conditioned to $\mu_f/a^2 > 10^6$, another way of expressing the result found by \cite{Lichtenberg2019}.
For their diagram on small body evolution, \cite{Moskovitz2011} adopted 1\,Pas for the melt viscosity and an ad hoc relationship between grain size and partial melting degree that inevitably boosts the melt migration: they considered an initial grain size of 100\,$\mu$m with an additional 90\,$\mu$m per percent of partial melting, such that the grain size reaches the millimeter scale at 10\% of partial melting. In this respect it is of note that the viscosity of a melt resulting from 10\% of partial melting of an H-chondrite material is in fact close to 1000\,Pas \citep{Dingwell2004} and decreases to 1\,Pas for 50\% of partial melting, the tipping point toward the magma ocean. 
In this respect, we note that the grain size in natural samples such as lodranites for which the degree of melting reached 20\% is around 500\,$\mu$m \citep{Keil2018}.
While the approach developed by \cite{Moskovitz2011} remains not only correct but also elegant, their evolutionary diagram should be reconsidered in the light of more recent knowledge on material properties such as grain growth laws and compositional and temperature effects on melt viscosity.

\cite{Lichtenberg2019} have also sketched the possible evolution of small bodies as a function of two parameters: the accretion time, and not their size but a dimensionless number, $\textrm{R}_{\textrm{seg}}$, which depends on almost all the parameters that control the physics of the differentiation except the body radius (fixed to 60\,km in their numerical models). 
$\textrm{R}_{\textrm{seg}}$ is a function of the ratio of two time scales, one related to the rate of heating by short-lived elements and the other corresponding to melt migration, or more precisely the Darcy flow rate of silicate liquid through the porous matrix that constitutes the unmelted residual rock.
At least two other time scales, both explicitly dependent on the radius, are involved in this problem:  
the first is the time during which the molten silicate persists, which is controlled by the cooling rate of the body, and the second is related to compaction of the matrix. 
Generally, the latter is considered to be non-limiting and therefore neglected, as in the model developed by \cite{Neumann2012}, because it is assumed to be too fast compared to the Darcy flow, except for very small bodies as we shall see.
The two time scales governing fluid migration, Darcy-flow and compaction, are affected, not only by temperature, but also by the thermal history itself via the size of the matrix grains whose growth over time is thermally activated.
Therefore, although the Darcy time scale is not explicitly expressed in terms of the size of the body, it also depends on it.

To describe the broad possible differentiation pathways for small bodies as a function of their time of accretion and size, we have chosen not to resort to a complete modelling of the thermochemical evolution as in \cite{Mizzon2015} or \cite{Lichtenberg2019}, but rather to follow the approach of \cite{Moskovitz2011} and to treat the thermal aspect and the migration of the fluids separately in order to extract the characteristic time scales from simple modeling. 


\section{Thermal time scales}
Various time scales can be defined to characterize the thermal evolution of small bodies, but some are more relevant than others for the present problem. Basically, three of them are of paramount importance: i) related to the heating rate, ii) related to the cooling rate, and iii) related to the life time of the heating sources, i.e. the short-lived radionuclides.

\cite{Lichtenberg2019} introduced the heating time scale as the ratio $\rho C_p \Delta T/ Q$, where $\rho$ is the density, $C_p$ the heat capacity, $Q$ the decay power per unit volume delivered by the short lived elements like $^{26}$Al and $^{60}$Fe which thus depends on the body accretion time (All notations and symbols used are listed in Table\,\ref{table:parameters}). 
$\Delta T$ is the temperature difference between the solidus and the accretion temperature, so that this time scale represents the time necessary to reach the solidus. 
This does not depend on the size of the body, despite the fact that a cooling rate that varies as the inverse square of the body radius is also involved. At equilibrium, the temperature of a body of radius $R$ heated by constant internal sources $Q$ reaches a maximum at its centre: $Q R^2/ 6 k_T$, with $k_T$ the thermal conductivity. For instance, at the time of the formation of the solar system, $Q$ is of the order of $6\times 10^{-4}$W/m$^3$ for H-type chondritic material, which implies a maximum temperature of 800\,K and 8\,K for bodies of 10\,km and 1\,km in diameter, respectively. This explains, for example, why dust, too small to retain heat, cannot be heated by the short-lived elements, and also why the elements $^{235,238}$U and $^{40}$K can heat planets, but not small bodies. Of course, these latter elements also heat planets because of their long lifespan. The life time of heat sources thus also plays a critical role. 

However, rather than using these three time scales (heating time, cooling time and radionuclide lifetime), we prefer to consider two time scales deduced from small body thermal evolution: a) the time necessary to reach the maximum melting degree from the onset of melting, and b) the time span of the existence of melt (i.e. the time interval spent above the solidus).  Both of these time-spans are affected by melt migration resulting in the fact that: 1) they cannot be determined analytically and 2) an approach based on dimensionless numbers (e.g. \cite{Lichtenberg2019}) cannot be used. Despite these limitations, relevant time-scales may be deduced from numerical simulations, and it is that approach that is taken here.

\subsection{Thermal evolution modelling}
The time necessary to reach the maximum melting degree from the beginning of melting and the time span of melt are derived from thermal evolutions of small bodies calculated from a model of pure conduction taking into account a realistic thermodynamic description of the melting of chondritic material. The model, previously used in a study devoted to the determination of the characteristics of the H chondrite parent body \citep[see][for numerical details]{Monnereau2013}, is a classical 1D spherical model based on the resolution of the conservation of heat. The latter can be written as:
\begin{equation}
\sum_{i=s,m} \rho_i \dfrac{\partial H_i}{\partial t} = \left[ \sum_{i=s,m}\rho_i \dfrac{d H_i}{dT}\right] \dfrac{\partial T}{\partial t} = \Div (k_T \Grad T) +\rho Q,
\label{eqn:EnergyC}
\end{equation}
with $H$ the enthalpy, $k_T$ the thermal conductivity and $Q$ the heat production by radioactive decay, subscipts $s$ and $m$ referring to the silicate and metallic components of the material that has been supposed to have a H chondrite composition \citep{Wasson1988}. The enthalpy of the silicate component is computed using a 1\,bar equilibrium melting simulation along the iron-wüstite (IW) 
buffer on the “Rhyolite-Melts" software \citep{Asimow1998, Ghiorso1995, Gualda2012, Ghiorso2015}. For small bodies with a radius less than a thousand kilometers, the energy is essentially provided by the decay of short-lived radionuclides such as $^{26}$Al. $^{60}$Fe is another short-lived radionuclide, with a comparable decay energy, but a half-life almost four times longer \citep{Castillo2009}. \cite{Quitte2010} showed that there were probably important heterogeneities of iron isotopes in the early solar nebula. The initial $^{60}$Fe/$^{56}$Fe ratio of the reservoir from which angrites and eucrites originated could be as low as $\sim  10^{-8}$ \citep{Quitte2011, Tang2012}. 
For some chondrites, it could be in the range of $4-7 \times 10^{-7}$ \citep[e.g.][]{Mishra2014}, which remains two orders of magnitude lower than the initial $^{26}$Al/$^{27}$Al ratio, so that we chose to neglect this radiogenic heat source. The heating power supplied by $^{26}$Al content is:
\begin{equation}
Q(t)=X_{\textrm{Al}}Q_0 \exp \left[ -\lambda_{^{26}\textrm{Al}} (t+t_{acc})\right],
\label{eqn:heat_sources}
\end{equation}
where $Q_0$ is the heating rate per mass unit of pure aluminium at CAI condensation, $t_{acc}$ the accretion time, $\lambda_{^{26}\textrm{Al}}$ the $^{26}$Al decay constant and $X_{\textrm{Al}}$ the aluminium mass fraction of the material. 

\subsection{Melting time scale and melt lifetime} \label{paragraph:thermal time scales}
\begin{figure}[t!]
\begin{center}
\includegraphics[width=8cm]{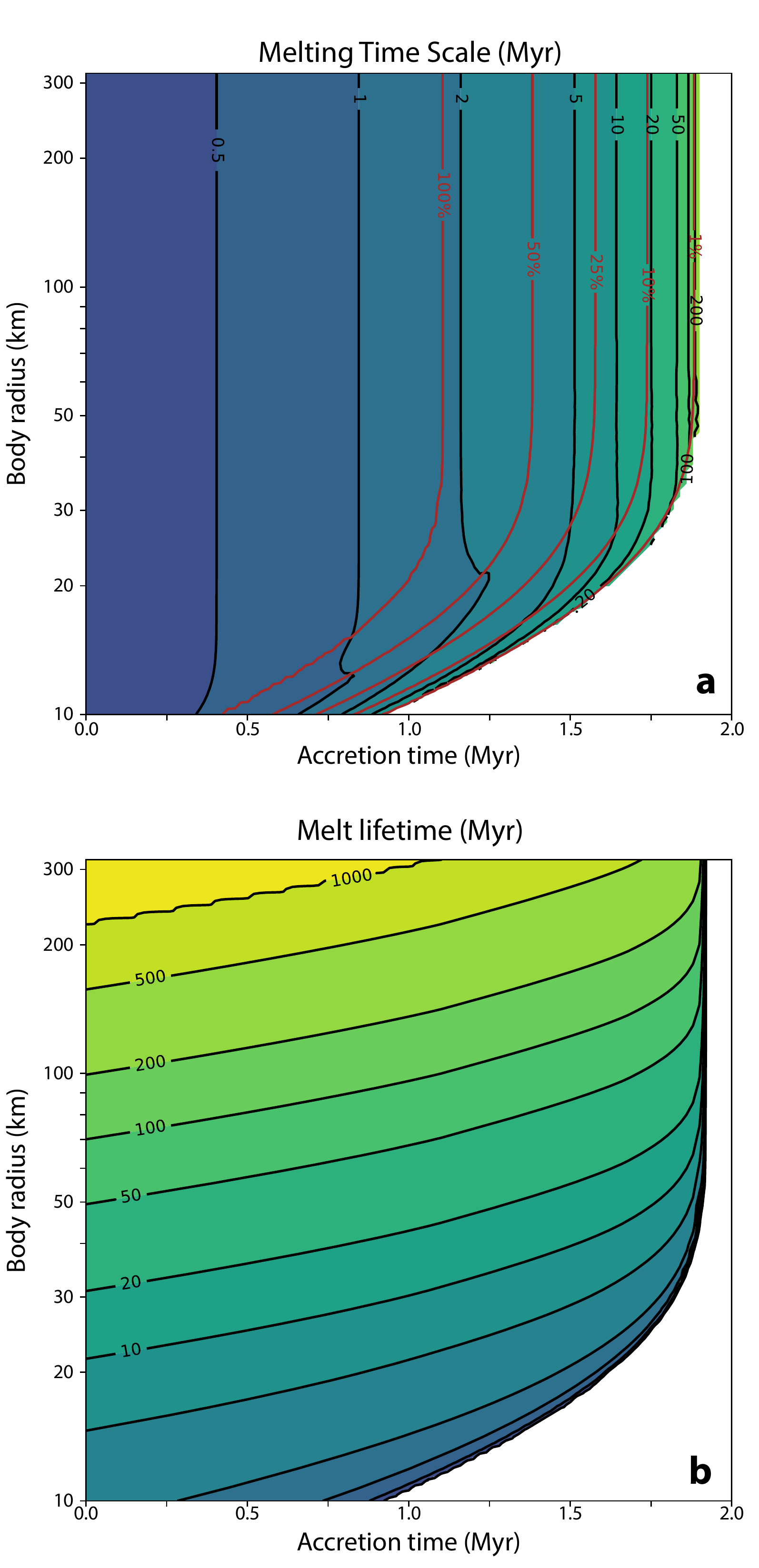}
\caption{The two time scales of thermal evolution of small bodies as a function of their accretion time and radius. (a) The melting time corresponds to the time elapsed between the beginning and the maximum of melting, or 50\% of melting. The time at the maximum of melting considered here is actually the time at the maximum of temperature minus 5 degree ; this choice permits to avoid the plateau characterizing the temperature evolution at the center of a body. Then, the 50\% melting threshold corresponds to the transition towards the development of a magma ocean beyond which silicate melt migration loses its relevance. (b) The melt time is the time during which at least 1 percent melt is present inside the body. These maps have been built from $101\times101$ numerical models of thermal evolution computed with 300 radial levels.}
\label{Fig:melting_times}
\end{center}
\end{figure}
The two thermal time scales relevant to melt migration, i.e. the melting rate time scale and the melt persistence time scale, are presented on Figure\,\ref{Fig:melting_times} as a function of the radius and accretion time of the body.

The first of these time-scales is related to heating power. Rather than considering the time required to reach the onset of silicate melting as in \cite{Lichtenberg2019}, the choice is made here to focus on the time span between the onset of melting and the maximum degree of melting. This choice is the result of the fact that we concentrate on the comparison of time intervals during which melt is present.
From a practical point of view, given that the temperature evolution at the centre is marked by a plateau before its maximum, we considered the time to reach  $T_\textrm{max}-5$K. 
The temperature to reach 50\% partial melting is another upper bound. As mentioned earlier, it marks the dissociation of the solid matrix and is accompanied by a drop in viscosity of several orders of magnitude. This is the tipping point for the melt migration to the magma ocean.
Therefore, Figure\,\ref{Fig:melting_times}.a shows the time required to reach $\min (T_\textrm{max}-5K, T_{50\%})$ since the onset of melting.

In this representation, we note that size only plays a role for bodies with a radius of less than 50 km, showing that the cooling time appears to be a limiting factor for melt production only for the smallest bodies. 
Above this size, the melting time is only a function of the age of accretion. Figure\,\ref{Fig:melting_times}.b displays the time during which at least 1\% of melting is present. 
As expected, this time increases as the square of the radius. It is also marked by a decrease when the accretion time increases. It should be noted that these times are deduced from the maximum temperatures reached at the centre of the body. They are therefore only representative of the main part of a body larger than 50 km, and only of a very restricted central part of smaller bodies.

\section{Melt drainage time scale}
The two thermal time scales presented above should be compared with a time scale related to the mobility of silicate liquids. For this purpose, we have chosen to calculate the time required for the separation of the fluid from the solid in an initial homogeneous mixture. These dynamics are described by the compaction equations, first introduced in geosciences by \cite{McKenzie1984}. Here, we used the formalism developed by \cite{Bercovici2001} and \cite{Bercovici2003} that treats both phases as equivalent. This formalism is summarized in \ref{paragraph:Two_phase_flow_formalism}.

\subsection{Drainage time for small bodies}

The rate of fluid-matrix separation depends on the ability of the matrix to deform and the fluid to flow through the matrix. The two processes have different characteristic times. 
The former, the compaction time $\tau_C$, can be defined as the ratio of matrix viscosity $\mu_m$ divided by the fluid-matrix difference pressure scale, $\delta \rho g_s R$: 
\begin{equation}
\tau_C = \dfrac{\mu_m}{\delta \rho g_s R} \propto \dfrac{\mu_m}{R^2},
\label{eqn:tauC}
\end{equation}
with  $\delta \rho = \rho_m - \rho_f$, the density contrast between the matrix and the fluid, R the radius of the body and $g_s$ the gravity at its surface. As $g_s$ is proportional to R, $\tau_C$ is inversely proportional to $R^2$.

The second characteristic time, the Darcy time $\tau_D$, is the ratio of the body radius $R$ to the filtration velocity: 
\begin{equation}
\tau_D = \dfrac{R \mu_f}{\delta \rho g_s k_0} \propto \dfrac{\mu_f}{a^2}.
\label{eqn:tauD}
\end{equation}
The filtration velocity is proportional to the fluid-matrix difference pressure, to the matrix permeability $k_0$ ---the permeability at the reference porosity $\phi_0$--- and inversely proportional to the fluid viscosity $\mu_f$. As a consequence, this time scale does not depend on the body size. Instead, the relevant length that appears here is the grain size, $a$. Indeed, the permeability is proportional to the square or the cube of the matrix grain size, depending on the pore geometry. Here, we adopted the classic permeability law considered for connected melt tubes :    $k(\phi) = a^2\phi^2/(72\pi)$ \citep{maaloe1982permeability}. Other geometries may be considered. For instance, in case of connected films, the permeability law becomes $a^3\phi^3/648$ \citep{Schmeling2000}.

Interestingly, the ratio of the two characteristic times corresponds to the square of the ratio of the two natural length scales of the problem, the body radius $R$ and the compaction length $L_c$: $\tau_D/\tau_C = R^2/L_c^2$. The compaction length is the length beyond which the compaction of a constant porosity matrix occurs \citep{McKenzie1984} and only depends on the properties of the matrix and the fluid: 
\begin{equation}
L_c=\sqrt{\dfrac{\mu_m k_0}{\mu_f}}.
\label{eqn:Lc}
\end{equation}

As mentioned above, since the surface gravity of a small body is proportional to its radius, it is also worth noting that the Darcy time does not depend on the size of the body, whereas the compaction time decreases as its square. As a consequence, in the case of porous flow dominated regimes, the radius of the body will not control the drainage characteristic time. On the contrary, when matrix deformation is the slower mechanism, the drainage time will be much longer for smaller bodies than for larger ones. 

\subsection{Drainage experiments}

\begin{figure}
\begin{center}
\includegraphics[width=7.2 cm]{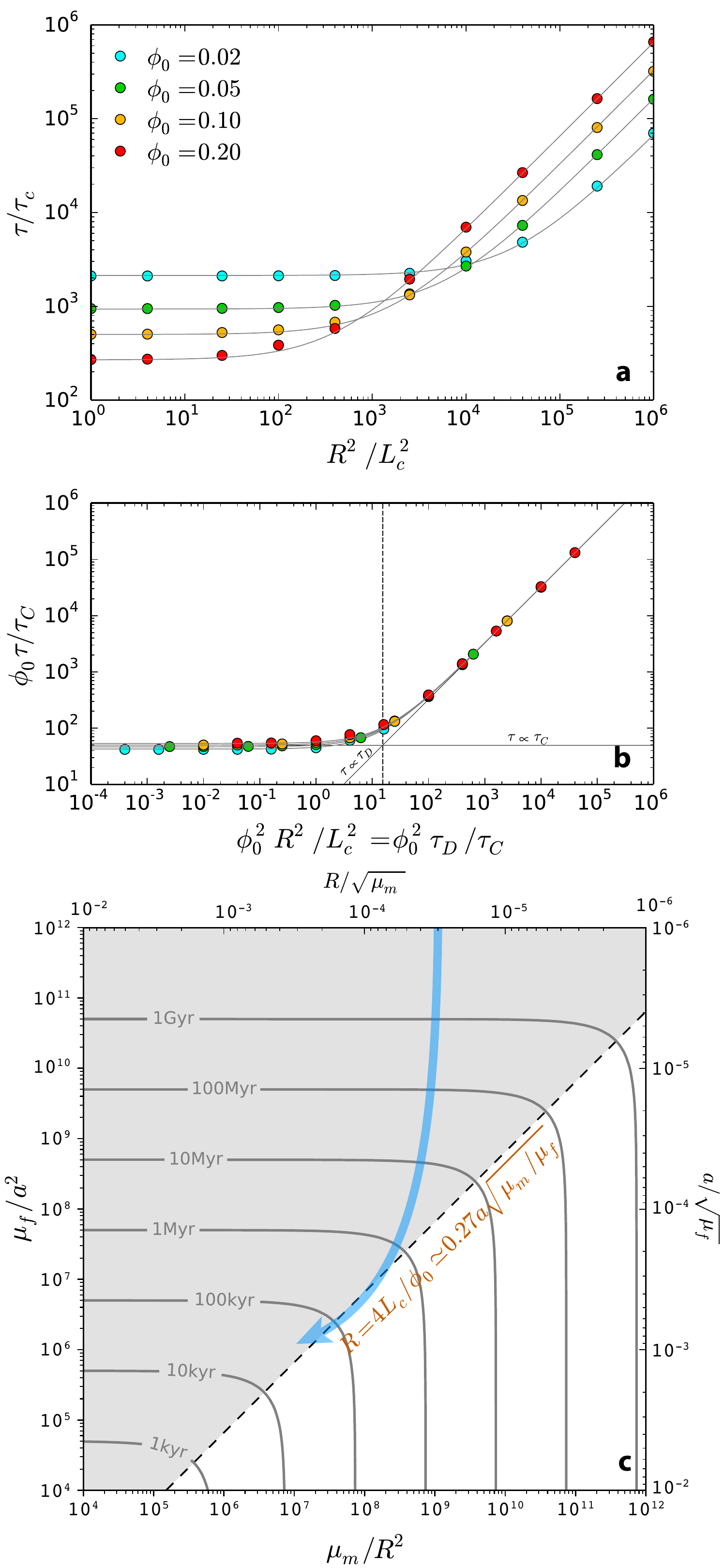}
\caption{Drainage time, $\tau$, i.e. the time required to reach 90\% of separation of the fluid from the matrix (see definition (\ref{eqn:segragation_1}) \& (\ref{eqn:segragation_2})) Four initial fractions of melt $\phi_0$ were tested, 2\% (cyan), 5\% (green), 10\% (yellow) and 20\% (red). (a) $\tau$, non-dimensionalized by the compaction time $\tau_c$ (\ref{eqn:tauC}), is plotted as a function of the non dimensional number $R^2/L_c^2$ where $L_c$ is the compaction length (\ref{eqn:Lc}). The solid lines correspond to the empirical fit $\tau = 63 \; \phi_0^{-0.9} \tau_C + 3.25 \; \phi_0 \tau_D$. This plot highlights the dependence of the drainage time on the initial fluid content $\phi_0$. (b) the quantity $\phi_0 \tau/\tau_c$ plotted as a function of $(\phi_0 R/L_c)^2$ appears almost independent on $\phi_0$. It is proportional to $\tau_c$ below $\phi_0 R/L_c \simeq 4$ and to $\tau_D$ above. (c) Plot of $\phi_0 \tau$ as a function of two independent parameters $\mu_m/R^2$ and $\mu_f/a^2$, $\phi_0 \tau$ being derived from the empirical relationship $\phi_0 \tau = 63 \; \phi_0^{0.1} \tau_C + 3.25 \, \phi_0^2 \tau_D$. The dash line, whose equation is $\phi_0 R/L_c=4$, separates regimes controlled by the compaction of the matrix (white) from regimes controlled by the Darcy flow (grey region). The latter is the one where compaction waves take place. The blue arrow describes the possible trajectory of a body in this parameter space due to progressive partial melting, which goes along with grain growth and decrease of both liquid and solid viscosities.}
\label{Fig:drainage_time}
\end{center}
\end{figure}

Drainage experiments consist in solving the flow equations (\ref{eqn:MassC_m}) \& (\ref{eqn:phi_adim}) with no source term, but from an initial constant porosity profile. They were performed for various values of $\phi_0$ and $R/L_c$ ratios. Figure\,\ref{Fig:drainage_time} reports the dependence of the drainage time $\tau$ as a function of both parameters. $\tau$ is arbitrarily defined as the time necessary to reach 90\% of segregation between the fluid and the solid. The segregation is evaluated through the function $s(\phi)$:
\begin{equation}
s(\phi)= 1 - \dfrac{1}{V \phi_0 (1-\phi_0)}\int \phi (1-\phi) dv,
\label{eqn:segragation_1}
\end{equation}
where $V=\int dv$ is the volume of the body. $s(\phi)$ is also the second central moment of the fluid distribution: 
\begin{equation}
\begin{split}
s(\phi)= \dfrac{1}{V \phi_0 (1-\phi_0)} \int (\phi-\phi_0)^2 dv.
\end{split}
\label{eqn:segragation_2}
\end{equation}
When the fluid is homogeneously distributed within the solid, its volume fraction is constant and equal to $\phi_0$ everywhere, $s(\phi_0)$ is then zero, while $s(\phi)$ is maximum and equal to 1 when the separation is complete, i.e. when $\phi$ is equal to 0 or 1. 

Figure\,\ref{Fig:drainage_time} shows that the drainage time is proportional to the compaction time for smaller bodies and to the Darcy time for larger ones. 
As explained above, in small bodies, the migration of fluid is controlled by the ability of gravitational forces to deform the matrix, whereas in large ones the limiting factor is the friction between the fluid and the matrix. 
Compaction waves develop in the latter case, while, in the former, the radial distribution of the liquid remains a monotonically increasing function of the radius. 
We observe the transition between both regimes around $\phi_0 R / L_c \sim 4$. An eye-fitting gives $\phi_0 \tau/\tau_C \simeq 63 \, \phi_0^{0.1} + 3.25 \, \phi_0^2 R^2 / L_c^2$. 
Since $\phi_0^{0.1}$ varies little for the range of porosity considered, the quantity $\phi_0 \tau/\tau_C$  essentially depends on the characteristic dimensionless number $\phi_0 R / L_c$. 
This is well illustrated by Figure\,\ref{Fig:drainage_time}b. 

The drainage time is found to be $\tau \simeq 63 \, \phi_0^{-0.9} \tau_C + 3.25 \, \phi_0 \tau_D$ and varies as function of three independent quantities $\phi_0$, $\mu_m / R^2$ and $\mu_f /a^2$ :
\begin{equation}
\begin{split}
\phi_0 \tau & \simeq \dfrac{3}{4 \pi \delta \rho \bar{\rho} G} \left[ 63 \, \phi_0^{0.1} \dfrac{\mu_m}{R^2} 
+ 3.25 \times 72 \pi \, \dfrac{\mu_f}{a^2} \right] \\
& \approx 4.3 \times 10^{4} \dfrac{\mu_m}{R^2} + 6.3 \times 10^{5} \dfrac{\mu_f}{a^2},
\end{split}
\label{eqn:empirical_phi_tau}
\end{equation}
with $\tau$, lengths and viscosities in SI. A graphical representation of this relationship is given in Figure\,\ref{Fig:drainage_time}c. The line $\phi_0 R/L_c=4$ separating the two regimes is an axis of symmetry for the quantity $\phi_0 \tau$. From any point located on this line, a decrease of $\mu_m/R^2$ or $\mu_f/a^2$ does not affect the drainage time, whilst an increase by a given factor increases the drainage time by the same factor. 
The blue arrow on Figure\,\ref{Fig:drainage_time}.c depicts what could be the trajectory of a body during its progressive melting, both matrix and fluid viscosity decreases and grains grow. A more precise description of such a trajectory is given in section \ref{sec:drainage_time_variation_during_melting} and the possible extent of the path in the parameter space is explained in the next section.

\subsection{The parameters of the drainage time}

\cite{Lichtenberg2019} emphasize the key role played by grain size in the differentiation of small bodies, pointing to inefficient percolation below a millimetre grain size. Here, we demonstrated that melt drainage is in fact controlled by two independent quantities, $\mu_m/R^2$ and $\mu_f/a^2$, in which the radius of the body, and the viscosities of the liquid and solid silicate are also involved.

\subsubsection{Grain size}
\label{paragraph:Grain_size}
During the thermal evolution of a body, the quantity $\mu_f/a^2$ is likely to experience a drop by six to seven orders of magnitude, jointly due to a decrease in liquid viscosity and an increase in grain size. 
First, the thermally activated silicate grain growth will promote a significant increase in grain size. As a matter of fact, the size of the silicate grains observed in meteorites increases with the maximum temperature they have experienced. It ranges from 10 to 100\,$\mu$m in chondrites depending on their metamorphic grade, goes up to 500-700\,$\mu$m in some achondrites like lodranites \citep{Krot2014} where the melting degree reached 20\%, and even to one centimeter for olivines in pallasites for which the melting degree has exceeded the threshold of the matrix disaggregation.

As grains grow more or less quickly depending on the temperature, their size can be computed by integration of the growth law along their thermal history. In general terms, normal grain growth is commonly described by an equation of the form:  \begin{equation}
a^n-a_0^n = A t,
\label{eqn:growth_law}
\end{equation}
where $a$ is the grain size at time $t$, $a_0$ the initial grain size. A is a thermally activated rate constant \citep{Atkinson1988}:
\begin{equation}
 A = A_0\,e^{-E_a/R_gT},
\label{eqn:growth_rate}
\end{equation}
with $E_a$ an activation energy, $R_g$ the gas constant and $A_0$ a constant. The exponent, $n$, theoretically an integer, may adopt various values from 2 to 5 depending on the mechanism controlling the grain coarsening \citep[e.g.][]{Brook1976,Atkinson1988, Evans2001}.

For the implementation of grain size evolution in their numerical model of small body differentiation, \cite{Neumann2012, Neumann2013, Neumann2014, Neumann2018} adopted a purely theoretical expression of relation (\ref{eqn:growth_rate}). They followed \cite{Taylor1993} who used the Lifshitz-Slyozov-Wagner relationship established for diffusion-controlled coarsening of particles in dilute solutions \citep[e.g.][]{Greenwood1969}. In this case the exponent is $n=3$ and the rate constant $A=8 \mathcal{V}^2 \gamma c D/9R_g T$, with $\mathcal{V}$ the molar volume of the silicate, $\gamma$ the surface free energy of the crystal-liquid interface, $c$ the equilibrium concentration of solute and D the diffusion coefficient. Beyond the fact that all these parameters are considered as constant in \cite{Taylor1993} and \cite{Neumann2012}, which yields a growth rate inversely proportional to temperature contrary to experimental observations, the assigned values (the same in both studies) lead to a rate constant $A=1.6\times 10^{-13}$m$^3$/yr at 1500K. This predicts a grain size of 0.55\,mm after $10^3$ yr, 1.2\,mm after $10^4$ yr, 2.5\,mm after $10^5$ yr and 5.4\,mm after $10^6$ yr; this latter value is higher by one order of magnitude than the size of the most evolved crystals measured in achondrites,  whose thermal history is well over a million years. The overestimation of crystal growth places the \citeauthor{Neumann2012}'s model well above the percolation efficiency threshold identified by \cite{Lichtenberg2019} and may affect the conclusions drawn in their papers. This will be discussed later.

 In addition to the many processes that may govern grain growth, various parameters can also be involved, such as the presence of impurities, secondary phases or melt. Therefore, only an experimental approach can identify those really at work. Several studies have been performed aimed at determining the values of the growth exponent $n$ and the activation energy $E_a$ on peridotitic material. 
 For the first one devoted to olivine, the theoretical growth exponent derived for a pure single phase,  n=2 or 3, ($E_a= 520, 600$ kJ/mol, respectively) has been satisfactorily fitted by experimental data run with San Carlos olivine at 0.1\,MPa and 300\,MPa in dry and wet condition \citep{Karato1989}. 
 However, in experiments conducted at 0.1\,MPa on dry aggregates of synthetic olivine Fo$_{91}$, \cite{Nichols1991} determined a value in the range of $n=4$ to 5 (with $E_a=290$\,kJ/mol and 345\,kJ/mol, respectively). They attribute this high value to the control of grain growth by coalescence through surface diffusion of a second phase, the residual porosity, $\sim 5\%$. 
 In sintered samples, porosity is inevitable in samples synthesized at 1 atm but almost disappears at high pressure, which may explain the low value of the exponent in the \cite{Karato1989} experiments at 300\,MPa, but not those at 0.1MPa. \cite{Nichols1991} noticed this disagreement without being able to explain it.
 Porosity has been measured in H-chondrites at equivalent levels and should have the same effect. Metallic grains, present in the chondritic material, also play the role of a secondary phase. \cite{Guignard2012, Guignard2016} studied a mixture of nickel and forsterite in proportions representative of the metal content of chondrites. 
 In this case, the growth of forsterite grains is limited  by that of the metallic particles whose growth process is controlled by diffusion along one dimensional paths due to their location at triple junctions, for which $n=5$. The activation energy was found to be close to 400\,kJ/mol. 
 Moreover, these experimental data were found to be consistent with the size of metal particles measured in H-chondrites at various metamorphic grades \citep{Guignard2016}, indicating that the parameters of this growth law are supported by natural points resulting from growth over several million years. 
 Lastly, in experiments on partially molten olivine aggregates conducted by \cite{Faul2006}, porosity cannot account for a pinning effect. However, the growth exponent $n$ is again measured above 3, close to 4 with an activation energy E=390kJ/mol similar to that measured for forsterite in the absence of melt. 
 Thus, in a single-mineral medium without impurities, the appearance of silicate liquid seems to reduce grain growth, while in poly-mineral or impurity-bearing medium, it has a promoting effect compared to conditions below the melting point.

For the present study, we used the relations (\ref{eqn:growth_law}) and (\ref{eqn:growth_rate}) with the parameters determined by \cite{Guignard2016} below the melting point and \cite{Faul2006} above, i.e. $n=5$, $E_a=400$\,kJ/mol, $A_0=10^{-19.04}$\,m$^5$/s and $n=4$, $E_a= 400$\,kJ/mol, $A_0= 10^{-12.02}$\,m$^4$/s, respectively. The initial grain size $a_0$ has been set in our calculations to 1 micron, a characteristic grain size of the matrix in primitive chondrites. 

\begin{figure}[t!]
\begin{center}
\includegraphics[width=7. cm]{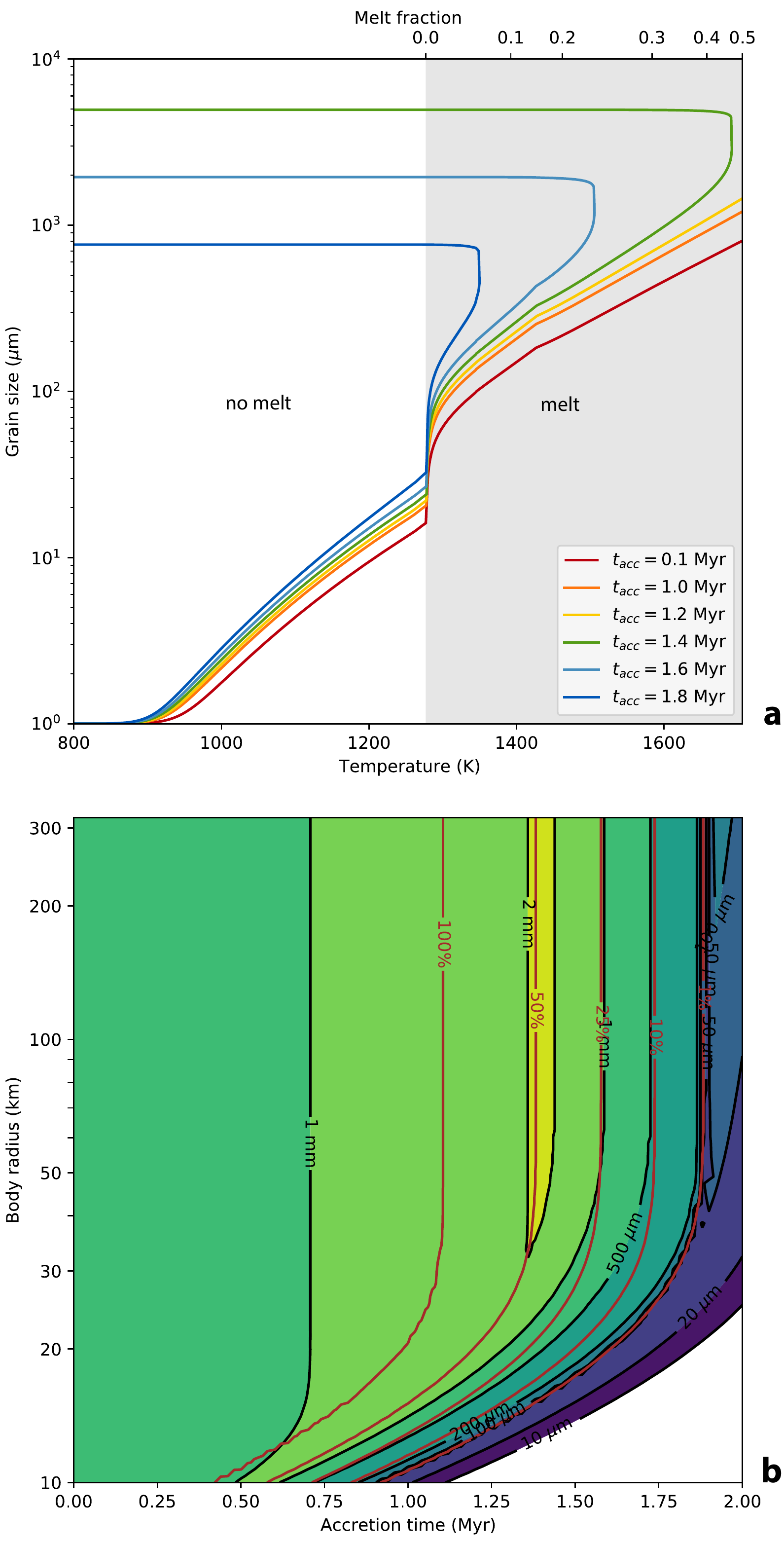}
\caption{Predicted size of silicate grains in the centre of a small body (a) over time for a body of 100km radius and different accretion times, (b) at $min(T_{max}-5K, T_{50\%})$. Grain growth laws used are from \cite{Guignard2016} below the melting point and from \cite{Faul2006} above. 
On panel (a), we note the increase in growth rate during melting, then the increase in size at constant temperature corresponding to the temperature plateau of the thermal evolution. 
The size reached at the end of this temperature plateau does not change during cooling, which corresponds to the horizontal part of each track. The curves have not been extended beyond $T_{50\%}$. 
On panel (b), we observe that the maximum grain size at $min(T_{max}-5K, T_{50\%})$ is reached for bodies that experience a maximum melting degree (red contour lines) of 50\%. In bodies accreted earlier, $T_{50\%}$ is reached more quickly, leaving less time for grain coarsening. In later accreted bodies, temperature is less important.}
\label{Fig:grain_growth}
\end{center}
\end{figure}

Figure\,\ref{Fig:grain_growth} displays the time integration of these growth laws along the thermal history of small bodies.
Figure\,\ref{Fig:grain_growth}a describes the time evolution of grain size located at the center of a 100km radius body accreted at various times. A kink marks the appearance of the liquid silicate and its influence on the growth of the grains, when the size increases from a few tens to a hundred microns with the first percent of liquid. 
After that point in time growth is significant at high temperature, notably during the long temperature plateau that follows the extinction of heat sources. 

Figure\,\ref{Fig:grain_growth}b shows a map, as a function of accretion time and body radius, of grain size reached just before this temperature plateau or before the tipping point to the magma ocean that occurs at 50\% of partial melting, i.e. at $\min (T_\textrm{max}-5K, T_{50\%})$ (see section \ref{paragraph:thermal time scales} for more detail on this threshold). 
On this map, the grain size reaches a maximum of 2\,mm for those bodies accreted at around 1.3\,million years, reaching a maximum of 50\% partial melting. 
For bodies formed earlier, the 50\% partial melting degree threshold is reached more quickly. This faster evolution does not allow the grain to grow as much, reducing the ability of the liquid to migrate during the melting phase. 

In summary, grain growth can account for a decrease of three or four orders of magnitude in the $\mu_f/a^2$ parameter.

\subsubsection{Viscosity of silicate liquid}
\label{paragraph:melt_viscosity}

\begin{figure} \begin{center} \includegraphics[width=8.5 cm]{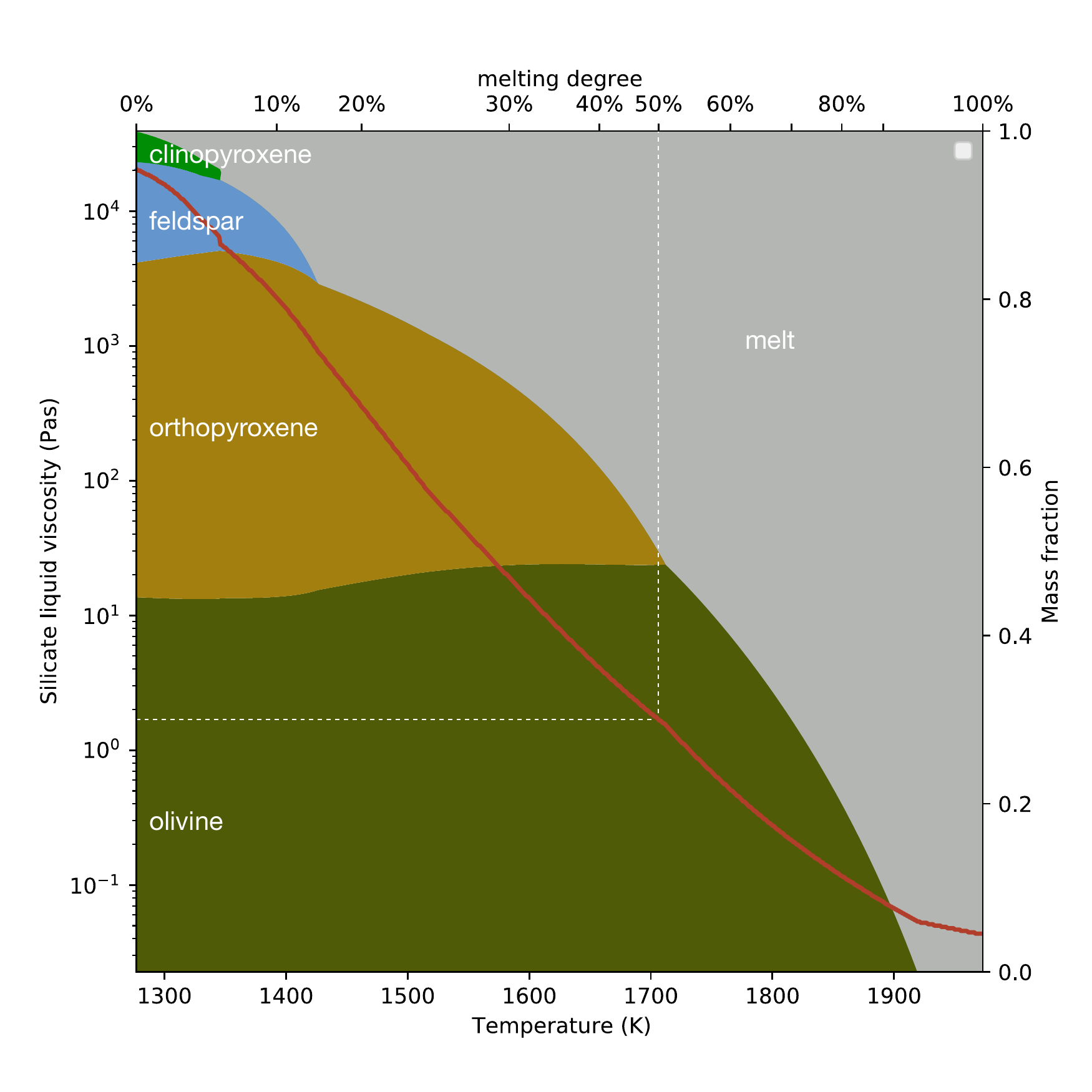} \caption{Viscosity of the silicate melt and composition of the residual solid as a function of temperature and melting degree (red curve). Composition and viscosity have been computed with the "Rhyolite-Melts" thermodynamical software \cite[][and references therein]{Ghiorso2015} from a H-type composition.} \label{Fig:melt_visco} \end{center} \end{figure}

The viscosity of liquid silicates varies strongly at the beginning of melting due to a dependence on the content of lattice elements, Si and Al, and notably Al that is only present in plagioclase, one of the earliest phases to melt-out of the matrix, with clinopyroxene. 
The first percentages of liquid have a viscosity of about $10^4$\,Pas \citep{Collinet2020}, which decreases to about 100\,Pas between 10\% and 15\% melt (basaltic liquids) and to about 1\,Pas above 40\% melt (picritic liquids) \citep{Dingwell2004}. Figure\,\ref{Fig:melt_visco} shows as an example the viscosity of a silicate liquid produced by melting of a H-chondrite composition as a function of temperature. 
This figure also shows the composition of the residual solid. 
Composition and viscosity of the liquid have been computed with the "Rhyolite-Melts" thermodynamic calculator \citep{ Ghiorso1995, Asimow1998, Gualda2012, Ghiorso2015}. 
We note that the reference value taken for the liquid viscosity by \cite{Moskovitz2011}, \citeauthor{Neumann2012} (\citeyear[][and the following papers]{Neumann2012})  and then \cite{Lichtenberg2019} corresponds to a liquid produced by about 50\% of partial melting (dashed line on Figure 4). 

\subsubsection{The parameter \texorpdfstring{$\mu_f/a^2$}{muf/a2}}
\label{paragraph:mu_over_a2}

\begin{figure}[t]
\begin{center}
\includegraphics[width=8.5 cm]{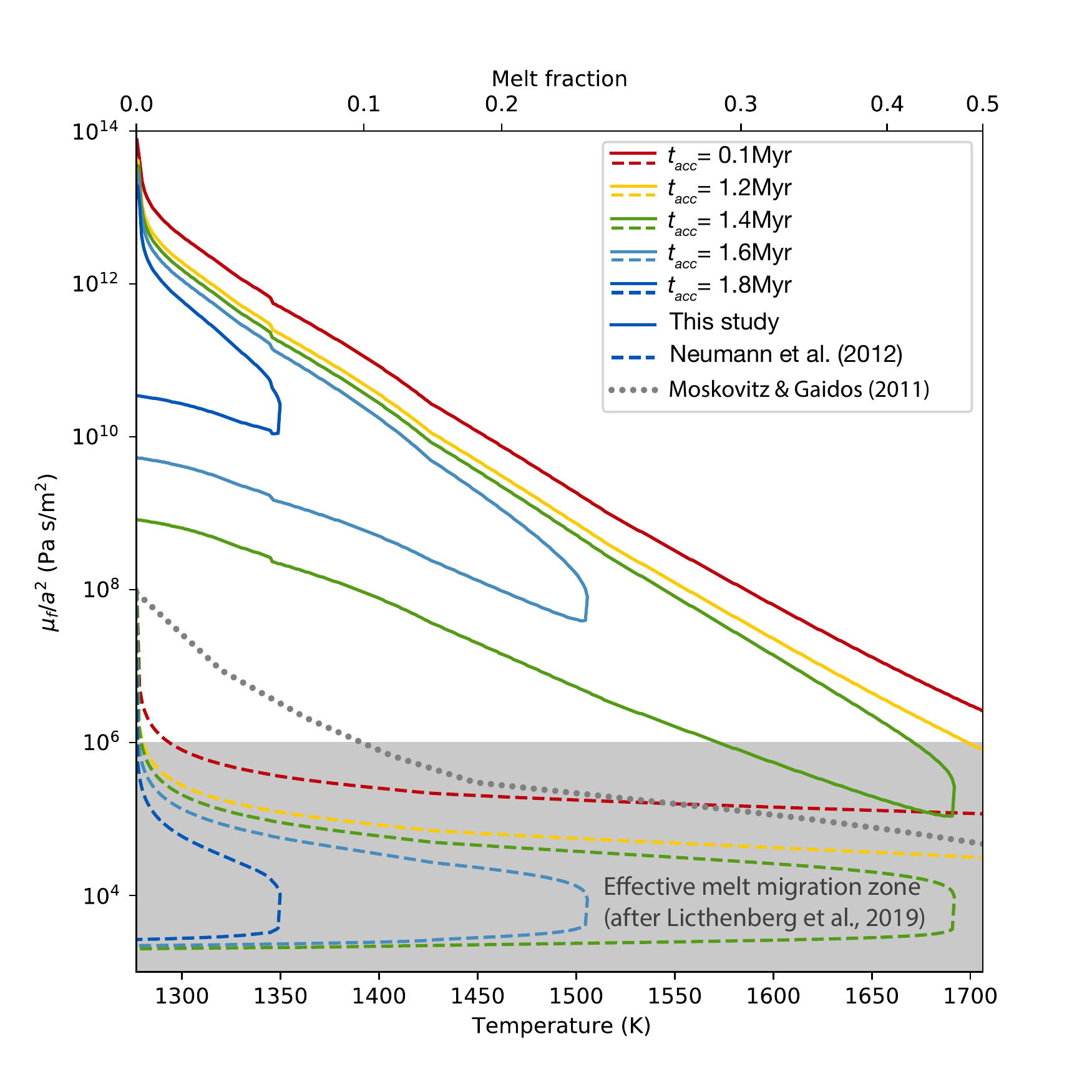}
\caption{The parameter $\mu_f/a^2$ calculated for the thermal evolution of a 100 km radius body accreted at different times, with the grain size and melt viscosity obtained from the laws chosen in this study (colored solid lines) and from grain growth laws used by \cite{Neumann2012} (colored dashed lines) and by \cite{Moskovitz2011} (grey dotted line), the melt viscosity being constant and fixed at 1 Pas in these last two studies. 
The grey shaded area corresponds to the domain where  melt migration is effective, i.e. where the grain size is larger than 1\,mm  \citep [after][since the melt viscosity is also set to 1\,Pas in this study]{Lichtenberg2019}.
In the latter two cases, the drainage time is 6 to 7 orders of magnitude lower than with realistic grain growth and melt viscosity laws, leading to an artificial drainage efficiency of silicate liquids and of heat sources well before their extinction}
\label{Fig:mu_f_over_a2}
\end{center}
\end{figure}
During melting, the variations of the different parameters can lead to a drop of seven to eight orders of magnitude in the $\mu_f/a^2$ ratio, which has a significant impact on the drainage time. For example, the blue arrow in Fig.\,\ref{Fig:drainage_time}c, shows the hypothetical trajectory of a body of radius 100\,km for which the drainage time is almost infinite at the beginning of the fusion ($a=10\,\mu$m, $\mu_m=10^{19}$\, Pas and the viscosity of the first liquids $\mu_f=10^{4}$\,Pas), to reach a state ($a=10^{-3}$\,m, $\mu_m=10^{17}$\,Pas and $\mu_f=1$\,Pas) corresponding to a draining time lower than 100\,kyr. Interestingly, the decrease of the $\mu_f/a^2$ ratio can lead to the region where the drainage is controlled by the compaction (white area on Figure\,\ref{Fig:drainage_time}c) and where its characteristic time then only depends on the viscosity of the matrix, whatever the grain growth or the melt viscosity decrease.

Figure\,\ref{Fig:mu_f_over_a2} presents the $\mu_f/a^2$ variations at the centre of a 100\,km radius body during its thermal evolution, computed following the grain growth and melt viscosity laws presented in Figures \ref{Fig:grain_growth} and \ref{Fig:melt_visco} respectively. 
It highlights this decrease of seven to eight orders of magnitude during melting for bodies accreted before 1.5 million years, and of lesser magnitude for later accretions. \cite{Lichtenberg2019} have shown that melt migration is inefficient for a grain size of less than 1\,mm. Since their calculations were performed with a constant melt viscosity of 1\,Pas,  this grain size threshold corresponds to a $\mu_f/a^2$ threshold of $10^6$ Pas/m$^2$. Importantly enough, figure\,\ref{Fig:mu_f_over_a2} shows that, except in some cases and for partial melt degrees above 40\%, the parameter $\mu_f/a^2$ remains above this value of $10^6$ Pas/m$^2$, i.e. in a range where melt migration is inefficient. 

For comparison, the $\mu_f/a^2$ parameter was also calculated with the grain growth laws used by \cite{Moskovitz2011} (grey dotted line) and by \cite{Neumann2012} (dashed lines), the melt viscosity being constant and fixed at 1\,Pas in both studies. 
Since the compaction of the matrix was neglected there, the drainage time is simply equal to the Darcy time and thus only proportional to the parameter $\mu_f/a^2$.
For both studies, the effective migration threshold of $10^6$ Pas/m$^2$ is crossed for the first percent of liquid, up to 10\% of liquid in the case of \cite{Moskovitz2011}. It is thus of no surprise that these studies have popularised the idea that silicate liquids extract rapidly to the surface taking the $^{26}$Al with them, leading to no further melting at depth. As an example, \cite{Moskovitz2011} showed that the percentage of liquid in the centre of a body formed at 1\,Myr would not exceed 27\%. 

However, the grain growth laws used in these two studies remain highly questionable. In \cite{Moskovitz2011}, it is an ad hoc dependence on the melt fraction, while the one used in \cite{Neumann2012}, if theoretically correct, adopts inappropriate parameters as discussed in paragraph \ref{paragraph:Grain_size}.
This, combined with the very low value chosen for the melt viscosity which does not correspond to the viscosity of the migrating melt, makes their conclusion about the high mobility of silicate liquids in small bodies questionable. 
On the other hand, the use of realistic laws for both grain growth and melt viscosity tends to show that liquids only start to migrate at melt fractions of several tens of percent, or even do not have time to extract before the rheological limit of a magma ocean is reached.

\subsubsection{Matrix viscosity}
\label{paragraph:Matrix_viscosity}

The drainage time is dependent on a second parameter, $\mu_m/R^2$, solely proportional to the matrix viscosity for a given body.
In this respect it is of note that the viscosity of solids depends on the creep mechanism at work. \cite{Lichtenberg2019} assume that the matrix deformation remains in the diffusion creep regime, but it could also be in the dislocation or grain boundary sliding (GBS) regimes, especially when the grain size is large and the stress is low. Here we will consider all of these possibilities, taking the most efficient one at each time step. 

\begin{figure*}[t!]
\begin{center}
\includegraphics[width=18. cm]{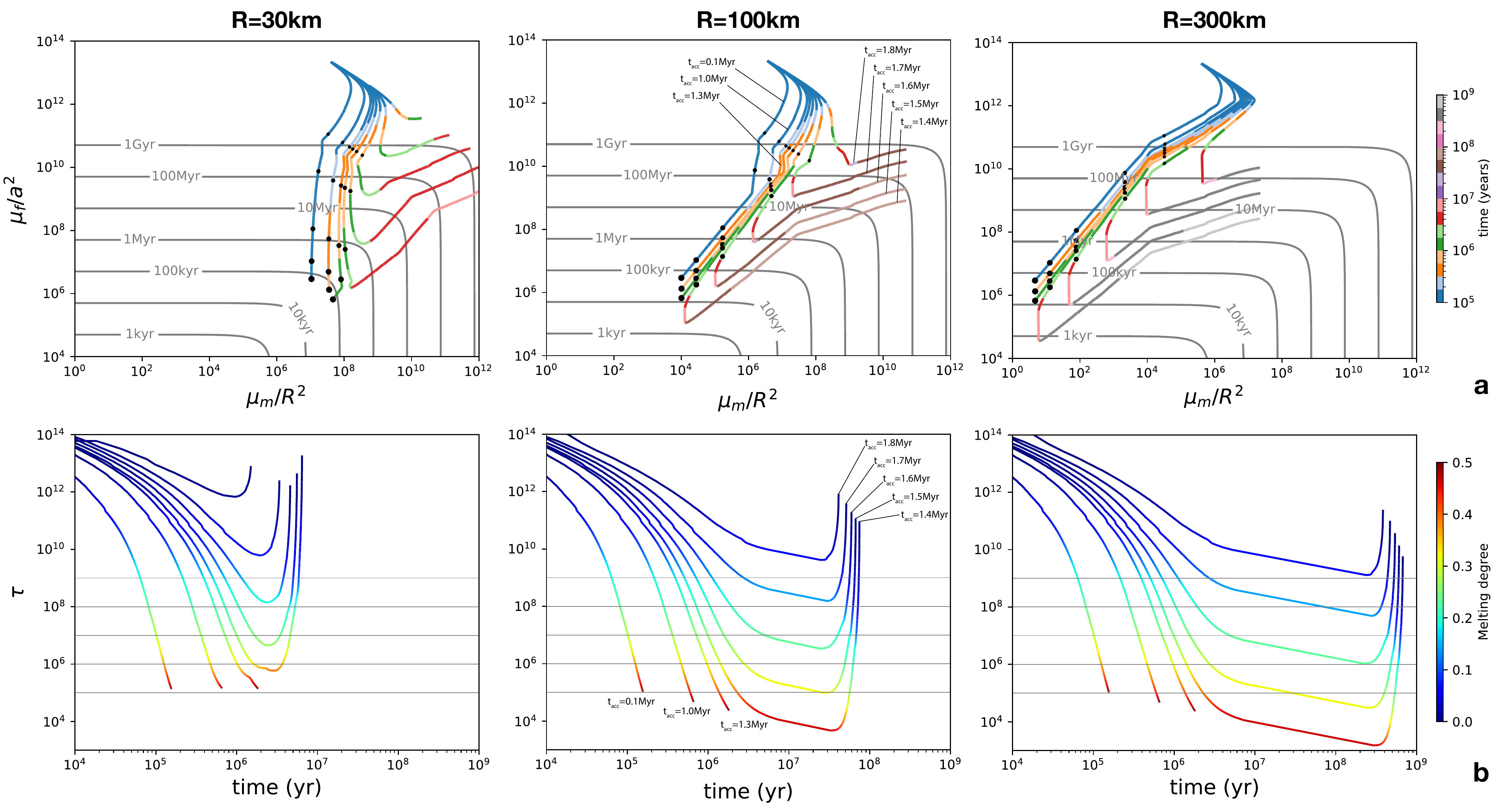}
\caption{(a) Time trajectories for three body radii and various accretion times drawn in $\mu_m/R^2$ and $\mu_f/a^2$ parameter space and on the corresponding $\phi \tau$ map. The color indicates the time since the onset of melting, tracks progressing from blue to grey.  The black dots mark the transitions to 10\%, 20\%, 30\%, 40\% and 50\% of partial melting. (b) Drainage time as a function of the time from the onset of melting. Color indicate the melting degree.
On both series of diagrams, trajectories are not plotted beyond the critical threshold of 50\% partial melting. }
\label{Fig:tracks_taudrain}
\end{center}
\end{figure*}

Whatever the creep mechanism, the rheological behavior of rocks is described by a general power law dependance of strain rate, $\dot{\epsilon}$, on differential stress, $\sigma$ \citep{Hirth2004}, which rewritten here in term of viscosity reads: 
\begin{equation}
\label{eqn:matrix_viscosity}
\mu_m = \dfrac{\sigma}{\dot{\epsilon}}=\dfrac{1}{A} \sigma^{1-q} a^n \exp(E_a/R_gT-\alpha \phi),
\end{equation}
where $A$ is a constant, $q$ the stress exponent, $n$ the grain size exponent, $E_a$ the activation energy and $a$ the grain size. $\alpha$ is a constant corresponding to the dependence on the melt content $\phi$. 
The presence of melt within an aggregate plays a well identified role on their deformation by favoring the sliding mechanism at grain boundary.
\cite{Hirth2004} reported a link between the effective viscosity and the melt fraction that can be approximated to first order by an exponential relationship up to liquid fraction of 12\%. It is not obvious that this law can be extended beyond that. Indeed, the viscosity measured in these experiments is the effective viscosity of the liquid-solid mixture. In the formalism we use, $\mu_m$ is the viscosity of the solid. The value of this parameter is subject to change due to the presence of liquid, notably because the presence of liquid promotes dislocations by accommodation at grain boundaries, resulting in a relaxation of the von Mises criterion. It nevertheless remains possible that this effect does not increase with increasing liquid fraction. For this reason, we will restrict the application of this law to the first 12 percent of liquid ( $\mu_m \propto \exp (-\alpha \min(\phi, 0.12))$), the range over which it has been determined experimentally. This choice has no incidence on our conclusions, as discussed in the next section. 

The coefficient $\alpha$ depends on the creep mechanism, estimated for lherzolitic material around 20 for diffusion creep and 25 for dislocation and GBS creeps. The other constants or exponents depend on the creep mechanism too. 
For instance, the stress dependence is linear for diffusion creep ($q=1$), but not for dislocation or GBS creep ($q=3.5$). 
In addition, there is no dependence on the grain size for dislocation creep, but it is nonlinear for diffusion ($n=3$) and for GBS ($n=2$) \citep{Hirth2004}.

The temperature, the liquid fraction and the grain size are calculated throughout the thermal history of the body, but here the stress can only be estimated \textit{a priori}. An upper estimate is the pressure difference at the centre of the body between the fluid and the matrix: 
\begin{equation}
\label{eq:stress}
\sigma= \dfrac{2 \pi}{3} \rho_m \delta\rho G R^2,
\end{equation}
Hence, $ \sigma \approx 1.41\times 10^{-3} R^2$ Pa, ranging from 0.5\,MPa to 50\,MPa for body radii from 30\,km to 300\,km, respectively.

The variation of all these parameters during the melting of silicates is likely to result in a variation of the matrix viscosity of six orders of magnitude, a range that is itself likely to shift by six orders of magnitude as a function of the size of the body, mainly because $\mu_m$ is proportional to $\sigma^{2.5}$ and finally to $ R^5$.

\subsection{Drainage time variation during melting}
\label{sec:drainage_time_variation_during_melting}

Figure\,\ref{Fig:tracks_taudrain}a displays the time trajectories for bodies of various sizes and accretion times on the $\phi_0\tau$ map shown on Figure\,\ref{Fig:drainage_time}c. These tracks highlight the magnitude of the variation of $\mu_m/R^2$ and $\mu_f/a^2$, the two independent parameters of the drainage time, and thus underline the importance of not considering material properties as constant in the study of small body differentiation. 

With the first percent of liquid, the trajectories begin with an increase in matrix viscosity and a decrease in the $\mu_f /a^2$ parameter, both due to the noticeable grain growth boosted by the appearance of silicate liquid, as shown in Figure\,\ref{Fig:grain_growth}a. Then, both parameters decrease while keeping the trajectories in the Darcy domain, such that the drainage time during melting is not very sensitive to variations in the viscosity of the matrix. This is particularly apparent for 100\,km and 300\,km radius bodies for which the variation in matrix viscosity differs greatly without affecting the drainage time. For the 30\,km bodies, this becomes less exact, their evolution pathways progressing into the compaction domain, where the variations in matrix viscosity regain their influence, i.e. where drainage time increases proportionally to the matrix viscosity.
Recall here that melt content has been limited to 12\% in the viscosity law.  Otherwise, we would observe a greater decrease in the matrix viscosity as melt content increases above this limit. However, this has no effect on the drainage time as long as the evolution remains in the Darcy domain. This would only affect the evolution corresponding to a 30\,km body accreted at 1.4\,Myr in its red part which could have progressed into the Darcy domain. This remains a secondary issue.
For each body size, the three trajectories corresponding to the earliest accretions reach the threshold of 50\% degree of fusion ---and have not been extended beyond that for this reason--- indicating a possible evolution into a global magma ocean. However, this end-point is not certain if the drainage time becomes shorter than the remaining time before reaching the threshold. This may be appreciated on Figure\,\ref{Fig:tracks_taudrain}.b that shows the time evolution of the drainage time since the onset of melting. In the case of very early accretion, within the first million years after CAIs and whatever the size of the body, the drainage time never becomes short enough to allow a significant drainage before reaching the threshold. 
In contrast, the trajectories at 1.3\,Myr show that the time required to reach the threshold remains greater than the drainage time after it falls below 1\,Myr. 
Thus, despite an accretion time that potentially allows the threshold to a magma ocean regime to be reached, migration of the liquid and heat sources to the surface is very likely, interrupting melting in the centre of the body and leading to shallower liquid accumulation that could itself develop into a shallow magma ocean. 
This scenario may also concern later accretion as in the case, for example, for 100\,km radius bodies formed between 1.3 and 1.5\,Myr, for which the drainage would be completed before the extinction of heat sources that occurs before 5\,Myr after CAIs. 
In that case, the concentration of heat sources may induce an overheating of the shallow liquid layer and a fusion of the overlying part \citep{Neumann2014, Lichtenberg2019}. 

Figure \ref{Fig:tracks_taudrain} also shows that differentiation may take place later, over the long cooling time of the body. This corresponds to accretions occurring between 1.5 and 1.7\,Myr, for bodies of 100\,km radius. Later accretions may lead to moderate partial melting but not to differentiation. The accretion time window for this regime is wider the smaller the body, as smaller bodies experience lower temperatures and faster cooling: between 1.5 and 1.8\,Myr for 30\,km radius against 1.7 to 1.8\,Myr for 300\,km radius.

\section{Types of evolution of small bodies }

In summary, both representations of Figure\,\ref{Fig:tracks_taudrain} help to distinguish various melt migration regimes. Three have already been described by \cite{Lichtenberg2019}: magma ocean, shallow sills and undifferentiated bodies. 
We add two more domains here by separating differentiation during melting with heat source transport from differentiation during cooling, and by subdividing undifferentiated bodies into those that have melted and those that have not.  
These 5 regimes correspond to conditions for which: 
\begin{enumerate}[label=\textit{\Roman*)}]
 \item the drainage is unable to prevent the melt fraction from reaching the 50\% rheological threshold and thus the development of a \textbf{global magma ocean} before differentiation;
 \item the drainage is efficient enough to allow extraction and subsurface accumulation of the melt and the heat sources it contains before their extinction, which may evolve into a \textbf{shallow magma ocean} above a residual harzburgitic or dunitic core;
 \item the drainage is not efficient enough to extract the melt before the heat source is extinguished, but efficient enough to allow a \textbf{moderate differentiation} of the body during its cooling; 
 \item \textbf{no differentiation} occurs despite moderate partial melting;
 \item there is no melting, just \textbf{thermal metamorphism}.
\end{enumerate}

\begin{figure} [t]\begin{center}\includegraphics[width=8. cm]{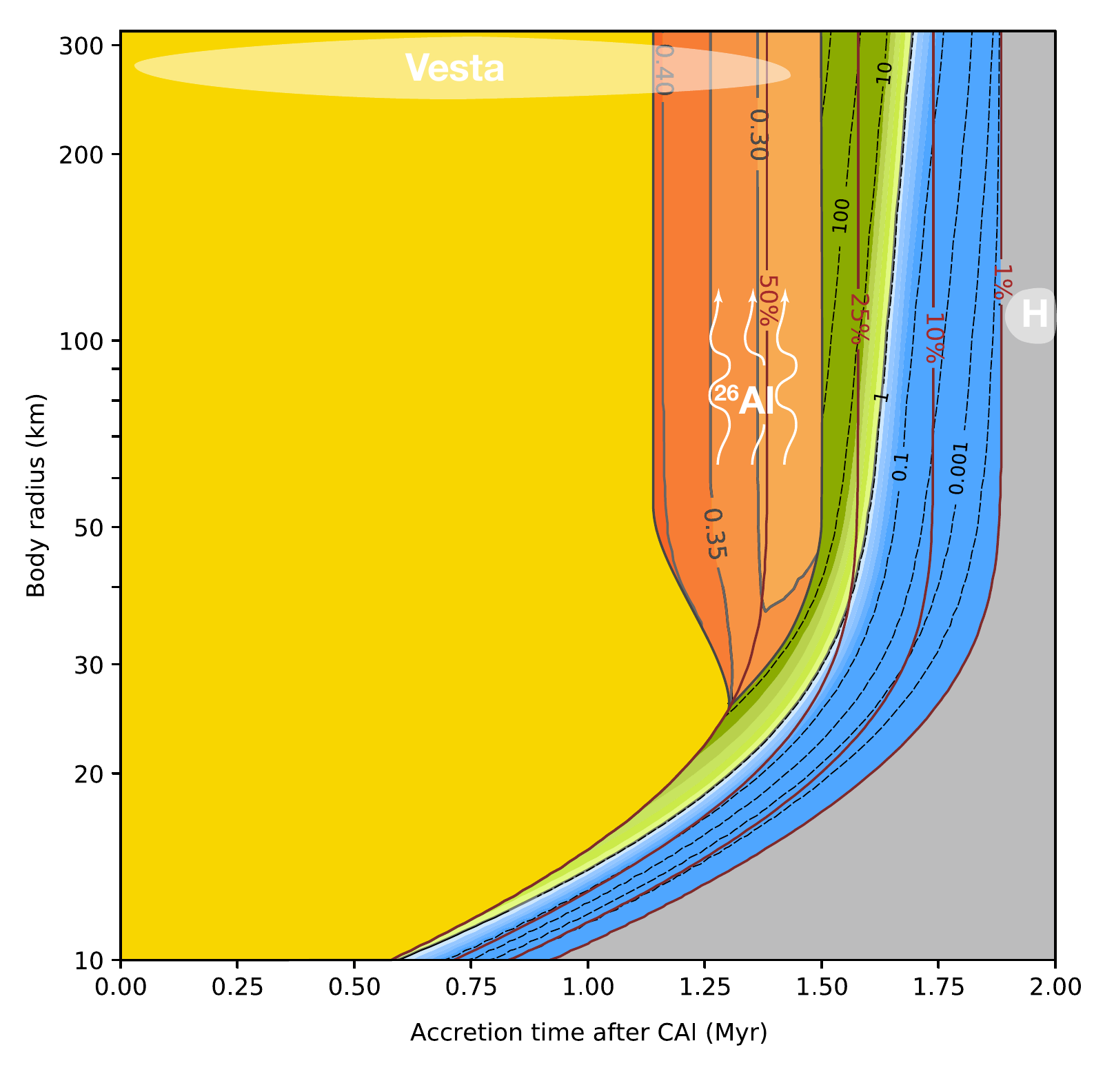}
\caption{Possible evolution for small bodies. \textit{I)}(yellow) Magma ocean: The internal heating is intense enough to push the interior of the body above the 50\% partial melting rheological threshold at which a magma ocean forms, before any liquid-solid segregation takes place; \textit{II)} (orange) Upward migration of the $^{26}$Al heat source: liquid-solid segregation is sufficiently efficient to drain the liquid before the extinction of the heat sources, leaving a 60\% to 70\% residual harzburgitic core; \textit{III)} (green) differentiation occurs after the extinction of heat sources, leaving a 70\% to 80\% harzburgitic core; \textit{IV)} (blue) interior partially melts up to 25\% without significant differentiation; \textit{V)} (grey) chondrite parent body domain.
Red contour lines indicate the maximum possible degree of partial melting. Black contour lines within the orange area indicate the degree of partial melting at which liquids become highly mobile and thus the degree actually reached after the heat sources have been removed. 
The dashed contour lines on the green and blue areas indicate the ratio of the melt lifetime to the drainage time. For instance, where this ratio is higher than 1 (green area), liquids have enough time to migrate and yield differentiation. Conversely, where it is lower than 1 (blue area) there is no migration and no differentiation.}
\label{Fig:big_picture}\end{center}\end{figure}

Figure\,\ref{Fig:big_picture} displays a global view of the different possible evolution pathways according to the radius and accretion time of the body. Criteria have been defined to delimit the five types. 
Type I (in yellow) corresponds to the appearance of a magma ocean. This must satisfy two criteria. Firstly, the maximum possible degree of partial melting, indicated by the red contours, must exceed the 50\% threshold. Secondly, this threshold must actually be reached. This second criterion defines the limit with type II. Type II corresponds to conditions that allow an efficient drainage before the end of melting or before reaching the 50\% melting degree threshold. To define this domain, the moment when the drainage time becomes smaller than the time remaining before reaching $T_\textrm{max} - 5\textrm{K}$ or 50\% of partial melting is inferred for each radius and accretion time based on Figure\,\ref{Fig:tracks_taudrain}.  
The orange region of Figure\,\ref{Fig:big_picture} represents the area where this criterion is satisfied. Inside, the contour lines indicate the melting degree at which the criterion is satisfied, showing that drainage only becomes effective from about 30\% of fusion. 
This percentage is not a threshold in itself, but results from several factors such as the viscosity of the liquid and the matrix, or the grain size. The migration of liquids certainly starts before this value, which is more indicative of the degree to which partial melting stops in the center of the body due to the drainage of heat sources. This value also constrains the composition of the unmelted residue remaining in the centre of the body. 

Outside the type II region, the 50\% partial melting threshold line separates type I, ending into a magma ocean, from types III (green) and IV (blue). The distinction between the latter two types is based on the ratio of melt lifetime, as defined in Figure\,\ref{Fig:melting_times}b, to the drainage time at its minimum value. On Figure\,\ref{Fig:big_picture}, the dashed contour lines plot the value of this ratio. Values less than one indicate an inefficient drainage at the time scale of the body cooling. To these four regions, we have added a fifth one (grey) corresponding to the conditions leading to non-melted bodies, i.e. the parent bodies of chondrites.

Naturally, these boundaries between regimes are more progressive than simple lines drawn in Fig.\ref{Fig:big_picture}, for two main reasons. 
On the one hand, as mentioned earlier, the drainage time is an average quantity of an active process depending strongly on evolving conditions. 
On the other hand, the criterion used to distinguish these types has been applied to conditions in the centre of the body that differs from those close to the surface. While these conditions may be representative of a large part of a body, a greater fraction as radius increases, the fact remains that its outer parts may have harbored conditions corresponding to another type. This is obvious for the parent bodies of chondrites, which may correspond to the outer parts of bodies of type III or IV.

This point is even more striking for very small bodies, those for which the radius is comparable to the thickness of the cooling boundary layer, thus those for which the central temperature is not representative of the body interior.
For example, the parent body proposed for the primitive achondrite Tafassasset, whose composition is compatible with a lost of basaltic content, has a radius possibly in the range of 15\,km to 30\,km with an accretion time between 0.5\,Myr  and 1.2\,Myr \citep[][]{Breton2015}. 
This falls in type I region, near the type III border. 
Details of the radial temperature profile derived from the thermal history modelling to satisfy isotopic dating shows a peak melting degree of 80\% in the center, while the overall composition of the meteorite ---olivine and pyroxene rich with plagioclase traces--- corresponds to a residue from 20\% of partial melting achieved at mid-radius, the outer 40\% of the radius remaining unmelted. 
Of course, such a description suffers from the lack of deep magma ocean and melt migration modelings and cannot be fully representative of the final evolution of a body accreted under such conditions.

\subsection{Did global magma oceans exist on small bodies?}

The occurrence of a global magma ocean in the early evolution of small bodies is not an idea that is unanimously accepted. 
Among the arguments against this hypothesis is the idea that liquid silicates drain efficiently towards the surface as soon as melting begins, either by filtration through the porous matrix resulting from partial melting \citep[e.g.][]{Moskovitz2011,Neumann2012}, or by percolation through a hierarchic network of small and large veins and dikes \citep[e.g.][]{Wilson2012, Wilson2017}.
However, this efficiency was deduced from a matrix permeability based on overestimated grain sizes and an underestimated viscosity of percolating liquids (see paragraphs \ref{paragraph:Grain_size} and  \ref{paragraph:melt_viscosity} for a detailed discussion), both leading to an overestimation of the percolation velocity by six to eight orders of magnitude compared to a calculation using realistic laws of crystalline growth or melt viscosity (see paragraph \ref{paragraph:mu_over_a2}).

In this respect, the theoretical arguments developed by \cite{Wilson2012, Wilson2017} deserve more detailed discussion. Briefly, these ideas are largely inspired by the context of mid-oceanic ridges. 
The formation of the veins and dikes network is conjectured from the matrix-liquid interplay described by the compaction formalism of \cite{McKenzie1984}, on which those exposed in \cite{Lichtenberg2019} or here in \ref{paragraph:Two_phase_flow_formalism} also rely \citep[see][for a detailed argumention of the vein network development]{Wilson2008}. 
To put numbers on his analytical developments, McKenzie uses a grain size of 1\,mm and a liquid viscosity of 1\,Pas. Interestingly, these values are those also employed by the other work supporting the ideas of efficient melt extraction \citep{Moskovitz2011, Neumann2012, Wilson2012, Wilson2017}, but they do not correspond to the conditions occurring at the beginning of melting on rocky-bodies in the early solar-system, where the grain size is closer to $100 \mu$m and the viscosity ranges around $10^4$\,Pas. 
As noted above, the values used in the literature are more relevant to conditions already close to the onset of a magma ocean, with the value of viscosity corresponding to a liquid produced by more than 40\% of partial melting. 

Hence, the work presented here indicates that extraction of a silicate melt from the interior of a small body is not as effective as believed up to now, and that the development of a global magma ocean is probable for a wide range of early accreted bodies. Indeed, most of the bodies accreted within the first 1.1\,Myr of CAI condensation probably experienced a global magma ocean stage.   

\subsection{The fate of magma ocean: pallasites, irons and eucrites}
\cite{Taylor1993} stated that \textit{“the lack of pyroxene in pallasites suggests that core formation in asteroids was accompanied by $>40\%$ melting of the associated silicate”.}
Indeed, pallasites provide simple but compelling evidence of a past magma ocean experienced by their parent body(ies). 
Aluminum is only present in feldspars, that have been completely removed from the solid residue after a partial melting degree of 15\%. If the silicate liquids are highly mobile and drain heat sources with them, melting cannot continue much further. 
Alternatively, if the liquids remain in the matrix, nothing can prevent the melting from reaching the rheological threshold, which incidentally corresponds to the disappearance of pyroxenes. 

With the onset of a magma ocean regime, thermal exchange switches to effective convective cooling. The temperature probably remains buffered at the rheological threshold, with the heat supplied by sources being used to melt and thin the layer above the ocean, up to the surface if sufficient, and then being dissipated outwards. 
While the demonstration of this point goes beyond the scope of the present study, metal-silicate separation in this context deserves discussion. 

First, metal is still present in the magma ocean even if it is sometimes assumed to be able to percolate through the silicate matrix as soon as it melts \citep[e.g.][]{Ghosh1998,Sahijpal2007, Neumann2012, Sramek2012}. 
To segregate, it should form a connected network along the silicate grain boundaries. 
Due to the high surface tension between the solid silicates and the liquid metal, this only occurs at a metal content of 20\,vol\%  \citep{Bagdassarov2009, Neri2020, Solferino2020}, well above the 10\,vol\% present in chondrites.
Thus, magma oceans inside planetesimals are a mixture of liquid silicate ($\sim$ 45\,vol\%), olivine crystals ($\sim$ 45\,vol\%) and liquid metal droplets ($\sim$ 10\,vol\%). 
This context differs from the that envisaged after giants impacts in which the metal-silicate separation is usually considered under the paradigm of iron rain in a fully liquid bath \citep{Rubie2003, Ichikawa2010}. 
Here, crystals are a significant component of the mixture and interact with the metal liquid. \cite{Neri2021} discussed this situation and showed that metallic drops remain attached to olivine crystals and probably sink as metal-olivine aggregates.
The separation finally occurs with the extraction of the interstitial silicate liquid during compaction of the matrix, leading to core formation. 
This process is similar to the scenario of pallasites formation where olivine crystals issued from 50\% of partial melting float in between the liquid core and the liquid silicate.

Iron meteorites are issued from parent bodies larger than 20km and accreted less than 1 Ma after the condensation of CAIs. They are the products of a high degree of partial melting and a complete metal-silicate differentiation; according to Figure\,7, their parent bodies indeed experienced a magma ocean stage.

Regarding asteroid Vesta, as it most likely has a metallic core \citep{Ermakov2013}, it must have formed within the first million years after the CAI's condensation. If it accreted later, the migration of liquids would have interrupted its melting and left a harzburgitic (olivine-pyroxene) center hampering metal separation. This conclusion issued from the present model is perfectly in line with cosmochemical data : indeed, based on isotope  geochronological data in eucrites, Vesta is known to have accreted very early \citep[e.g.][]{Schiller2011}.

\subsection{Incipient differentiation in the absence of a magma ocean}
Some other meteorites such as primitive achondrites (winonaites, acapulcoites, lodranites) are characterized by a lower degree of partial melting and incipient differentiation. In more detail, winonaites are thought to have been heated slightly above the Fe-FeS eutectic temperature and potentially slightly above the silicate solidus \citep[e.g.][]{Hunt2017}, acapulcoites show signs of extraction of small amounts of silicate and Fe-Ni-S melts, while lodranites point to a maximum degree of $\sim 20\%$ partial melting and the removal of both a basaltic and a S-rich metallic melts \citep{Mccoy1997}. In all these cases, the melting degree does not reach the threshold required for the onset of a magma ocean. 
Even so, winonaites, lodranites and accapulcoites might not be representative of the center of their parent bodies, but of the outer parts, as proposed by \cite{Neumann2018}.
Combining data inferred from different chronometers together with modeling, they concluded that acapulcoites and lodranites could be issued from a single parent body 260\,km in radius, accreted at 1.7\,Myr after CAIs with burial depth of $\approx$ 7 to 13\,km for acapulcoites and lodranites respectively. 
Interestingly, they noted that their model is unable to fit the thermo-chronological data with the grain coarsening law and the 1\,Pas melt viscosity that they usually adopt, because the migration of the silicate melt along with the enrichment of $^{26}$Al disturbs the evolution of the temperature. They found 
a suitable fit, either by inhibiting the partitioning of $^{26}$Al heat sources between matrix and silicate melt, or by keeping small grain size (0.2\,$\mu$m), or by increasing the melt viscosity up to 100\,Pas. 
We note that a 260\,km radius body formed at 1.7\,Myr lies at the boundary between  the green and blue areas of Fig.\ref{Fig:big_picture}, where silicate melt migration is possible, but to slow to allow heat sources redistribution. These regions correspond to regime III and IV, which are compatible with lodranites and acapulcoites resppectively.

\section{Key summary points }

Contrary to the idea popularised by \cite{Moskovitz2011} and the series of articles published in the following decade, we show here that melt migration is not a process efficient enough to drain and remove $^{26}$Al heat sources from the interiors of small bodies before their heat sources are exhausted. 
This conclusion is drawn from new considerations regarding grain size and melt viscosity. 

In summary, melt percolation rate is proportional to the ratio of matrix permeability to melt viscosity, i.e. the ratio of the square of the matrix grain size to the melt viscosity. 
Literature studies used ad hoc or unsuitable grain growth laws --- that overestimate grain size by more than one order of magnitude---, associated to the canonical melt viscosity value, 1\,Pas, chosen by \cite{McKenzie1984} to assess melt migration rates in the context of MORB generation at mid-oceanic spreading centres.
Up to a partial melting degree of 20\%, the relevant values are closer to  100\,$\mu$m and $10^3$\,Pas, corresponding to a migration rate five order of magnitude lower. 

This finding underlines the importance of not assuming constant values for properties that play a role in magma migration, when modelling the differentiation of small bodies. Other consequences concerning small body evolution are:
\begin{enumerate}
    \item Small bodies accreted within 1.15 million years of CAIs condensation underwent a magma ocean stage which permitted the formation of a metallic core. These bodies are possible parent bodies of pallasites; 
    \item Melt migration and drainage of $^{26}$Al heat sources influenced the differentiation of bodies larger than 30km radius and accreted between 1.15 and 1.5\,Myr after CAIs, leaving an olivine-pyroxene rich core overlain by a shallow magma ocean;
    \item The parent bodies of winonaites, accapulcoites, and lodranites are consistent with expectations for bodies accreted more than 1.5\,Myr after CAIs condensation.
\end{enumerate}

\section*{Aknowledgments}
This work is part of the PALLAS project funded by the ANR (grant ANR-14-CE33-006-01 to G. Quitté); we also thank the Université Paul Sabatier for its contribution to the PhD grant to A. Néri.

\appendix

\section{Two phase flow formalism} \setcounter{figure}{0} 
\label{paragraph:Two_phase_flow_formalism}
To describe the migration of the silicate melt within a small body, we follow \cite{Sramek2012} that extended the two-phase flow formalism of \cite{Bercovici2001} and \cite{Bercovici2003} in spherical geometry. Here, we just recall the principal steps of this approach. This formalism consists in averaging on a unit volume the mass and conservation equations written for the fluid and the matrix. This introduces an additional variable that describes the volume fraction of the fluid $\phi$, which will require a closure relation to be solved \citep{Drew1998}. 

\subsection{Conservation equations}

\begin{table*}[t!]
\caption{Notations and parameters.}
\small
\begin{tabular*}{\hsize}{@{\extracolsep{\fill}}llll}
\hline
 Symbol & Quantity & value & unit \\
\hline
$\mathcal{A}$ & Avogadro's number & $6.02214076 \times 10^{23}$&atom/mol\\
$E_{^{26}\textrm{Al}}$ & $^{26}$Al decay energy$^a$ & 3.12 &MeV/atom\\
$\lambda_{^{26}\textrm{Al}}$ & $^{26}$Al decay constant & $3.063\times10^{-14}$ &
s$^{-1}$\\
$m_{^{26}\textrm{Al}}$ & $^{26}$Al molar mass & 0.026 & kg/mol\\
$X_{\textrm{Al}}$ & Mass fraction of Al in H-type chondrites$^b$ & $11.3\times10^{-3}$ & \\
$[^{26}\textrm{Al}/^{27}\textrm{Al}]^0$ & initial $^{26}\textrm{Al}/^{27}\textrm{Al}$ ratio$^c$ &$5 \times 10^{-5}$&\\ 
$H$ & enthalpy & & J/kg \\
$k_T$ & thermal conductivity$^d$ & $4\sqrt{Te/T}$ & W/m/K \\
$Q_0$ & initial radiogenic heat power per Al kg & $\left[^{26}\textrm{Al}/^{27}\textrm{Al}\right]^0 \mathcal{A} \lambda_{^{26}\textrm{Al}} E_{^{26}\textrm{Al}} / m_{^{26}\textrm{Al}}\simeq 1.77\times10^{-5}$ & W/kg \\
$Q$ & radiogenic heat power & & W/kg\\
$T$ & temperature& &K\\
$T_e$ &external temperature temperature & $292$ & K\\
$a$ & grain size & & m \\
$a_0$ & initial grain size$^{e}$ & $10^{-6}$ & m \\
$A$ & grain growth rate & & \\
$A_0$ & growth rate constant & & \\
$\vect{g}$, $g_s$ & gravity, surface gravity& & m/s$^2$\\
$E_a$ & activation energy & & J/mol\\
$G$ & gravitational constant& $6.67430 \times 10^{-11}$& m$^3$/kg/s$^2$\\
$k$ & permeability & &m$^2$\\
$k_0$ & reference permeability & &m$^2$\\
$K$ & non dimensional geometrical factor & &\\
$L_c$ & compaction length & &m \\
$P_{f, m}$ & fluid, matrix pressure & & Pa\\
$r$ & radius & & m\\
$R$ & surface radius & & m \\
$R_g$ & gaz constant & 8.314 & J/mole/K \\
$s(\phi)$ & segregation function & & \\
$t$ & time & & s\\
$\vect{v}_{f, m}$ & fluid, matrix velocity& & m/s\\
$\mu_{f, m}$ & fluid, matrix viscosity$^f$,$^g$ &  & Pas \\
$\rho$ & average chondritic material density$^h$ & 3800 & kg/m$^3$ \\
$\delta\rho$ & solid-liquid density contrast & 1000 & kg/m$^3$ \\
$\tens{\sigma}_m$ & matrix deviatoric stress tensor & & Pa \\
$\tau_D$ & Darcy characteristic time & & \\
$\tau_C$ & compaction characteristic time & & \\
$\phi$ & fluid volume fraction, matrix porosity& 	&\\
\hline
\end{tabular*}
$^a$\cite{Castillo2009}, $^b$\cite{Wasson1988}, $^c$\cite{Macpherson1995}, $^d$\cite{Monnereau2013},  $^{e}$\cite{Krot2014}, $^f$\cite{Dingwell2004}, $^g$\cite{Hirth2004},$^h$\cite{Consolmagno2008} \\
\label{table:parameters}
\end{table*}
In absence of phase change, the formulation of the mass conservation for the fluid and the matrix are standard (hereinafter, subscripts $f$ and $m$ refers to the fluid and the matrix phases respectively; all notations are summarized in Table\ref{table:parameters}.):
\begin{equation}
\dfrac{\partial \phi}{\partial t} + \Div \left[\phi \vect{v}_{f} \right] = 0
\label{eqn:MassC_f}
\end{equation}
and
\begin{equation}
-\dfrac{\partial \phi}{\partial t} + \Div \left[(1-\phi) \vect{v}_{m} \right] = 0
\label{eqn:MassC_m}
\end{equation}
whose sum shows that the average velocity $\bar{\vect{v}} = \phi \vect{v}_f +(1-\phi)\vect{v}_m$ is solenoid: 
\begin{equation}
\Div \bar{\vect{v}} = 0.
\label{eqn:Div_v}
\end{equation}
The average and difference quantities of the phases are defined as $\bar{q}=\phi q_f + (1-\phi) q_m$ and $\Delta q = q_m -q_f$. If surface tensions are neglected, the conservation of momentum for the fluid and the matrix phases can be written as \citep[see][for details]{Bercovici2003}:
\begin{equation}
-\phi \left[ \Grad P_f - \rho_f \vect{g} \right]+ c\Delta\vect{v}=0
\label{eqn:MomC_f}
\end{equation}
and 
\begin{equation}
-(1-\phi) \left[ \Grad P_m - \rho_m \vect{g} \right]+ \Div \left[ (1-\phi) \tens{\sigma}_m\right] - c\Delta\vect{v}=0.
\label{eqn:MomC_m}
\end{equation}
$P$ is the pressure. The term $\Div \left[ (1-\phi) \tens{\sigma}_m\right] $ is the viscous dissipation, where $\tens{\sigma}_m$, the matrix deviatoric stress tensor, is:
\begin{equation}
\tens{\sigma}_m = \mu_m\left[ \Grad \vect{v}_m + \left[ \Grad \vect{v}_m \right]^\intercal -\dfrac{2}{3}(\Div \vect{v}_m)\tens{I} \right],
\end{equation}
$\mu_m$ being the matrix viscosity. Because of the very low fluid viscosity compared to the matrix viscosity ($\mu_f \ll \mu_m$), this dissipation term is neglected in the fluid equation. The term $c\Delta \vect{v}$ is the Darcy term where $c$ is the drag coefficient between both phases that, in case of large viscosity ratio between the matrix and the fluid, reduces to \citep{Bercovici2003}:
\begin{equation}
c = \dfrac{\mu_f \phi^2}{k(\phi)}.
\label{eqn:c_def}
\end{equation}
$k(\phi)$ is the permeability. The term $\Delta P \Grad \phi$, appearing in the matrix momentum conservation (\ref{eqn:MomC_m}), is the interfacial pressure force acting between both phases. It appears in each momentum equation with a weighting coefficient that is zero for the fluid momentum equation when $\mu_f \ll \mu_m$.

The mass conservation equations (\ref{eqn:MassC_f}) \& (\ref{eqn:MassC_m}) and momentum conservation equations (\ref{eqn:MomC_f}) \& (\ref{eqn:MomC_m}) form an incomplete system to solve the five unknowns ($P_f, P_m, \vect{v}_f, \vect{v}_m$ and $\phi$). The conservation of energy and damage allows to write a closure relation that, under the previous approximations, is \citep{Bercovici2001}:
\begin{equation}
\Delta P= -K \dfrac{\mu_m}{\phi} \Div \vect{v}_m,
\label{eqn:Closure}
\end{equation}
where $K$ is a dimensionless factor related to the geometry of the two-phase mixture, taken equal to 1 in \cite{Bercovici2001}. The combination of momentum equations, $(1-\phi)$(\ref{eqn:MomC_f}) - $\phi$ (\ref{eqn:MomC_m}), yields to the action-reaction equation: 
\begin{equation}
\begin{split}
-\Grad \left[ (1-\phi) \Delta P \right] &+ (1-\phi) \Delta \rho \vect{g} \\
&+ \Div \left[ (1-\phi) \tens{\sigma}_m \right] - \dfrac{c \Delta \vect{v}}{\phi} = 0
\end{split}
\label{eqn:AR}
\end{equation}
that gives an equation for the matrix velocity after substitution of $\Delta P$ from (\ref{eqn:Closure}) and of $\Delta \vect{v}$ from (\ref{eqn:Div_v}). Equation (\ref{eqn:Div_v}) indicates that the average velocity is constant. In spherical geometry, it is necessarily null, so that:
\begin{equation}
\vect{v}_m = \phi \Delta \vect{v}.
\label{eqn:matrix_velocity}
\end{equation}

In 1D spherical geometry, the equation for the matrix velocity is thus (the full development of these equations in spherical geometry can be found in \cite{Sramek2012} or in \cite{Mizzon2015}):
\begin{equation}
\begin{split}
&k(\phi) \frac{\partial}{\partial r} \frac{(1 - \phi)}{r^2} \left[ \dfrac{1}{\phi} + \dfrac{4}{3}\right] \frac{ \partial r^2 v_m}{\partial r} \\
&+\left[ \dfrac{4 k({\phi})}{r} \dfrac{\partial \phi}{\partial r} -\frac{\mu_f}{\mu_m}\right] v_m = \dfrac{k(\phi) (\bar{\rho} - \rho_f)g(r)} {\mu_m}.
\end{split}
\label{eqn:final_phi}
\end{equation}
$g(r)$, the radial gravity profile, is obtain by integration of the average density profile, $\bar{\rho}(r)$ which is assumed to be only dependent on the composition ---due to the size of small bodies, the pressure dependency is neglected---:
\begin{equation}
\label{eqn:gravity}
g(r)=\dfrac{4\pi G}{r^2}\int_0^r \bar{\rho}(x) x^2 dx.
\end{equation}
Solving equations (\ref{eqn:MassC_m}) and (\ref{eqn:final_phi}) allows the calculation of the liquid and solid fraction as a function of depth and time. This is achieved through a finite volume discretization and a second order Runge-Kutta method. More details can be found in \cite{Mizzon2015}.

Choosing $R$ and $\tau_{C}$ (cf equation \ref{eqn:tauC}) as length and time scales, the matrix velocity equation (\ref{eqn:final_phi}) reads:
\begin{equation}
\begin{split}
 \frac{\partial}{\partial r} \frac{F}{r^2} \frac{ \partial r^2 v_m}{\partial r} + \left[ \dfrac{4 }{r}\dfrac{\partial \phi}{\partial r} - \frac{\phi_0^2 R^2}{\phi^2 L_c^2} \right] v_m 
= (1-\phi)g
\end{split}
\label{eqn:phi_adim}
\end{equation}
where $F=(1-\phi)\left[1/\phi +4/3\right]$. $r$, $v_m$ and $g$ are dimensionless variables. $g$ is the gravity acceleration normalised by the surface gravity $g_s$. A dimensionless number appears in this equation: 
\begin{equation}
\dfrac{\phi_0^2 R^2} {L_c^2} =\dfrac {72 \pi R^2 \mu_f} {a^2 \mu_m}.
\end{equation}
It is worth noting that it does not depend on $\phi_0$.

\subsection{Analytical solutions}
\begin{figure}[t]
\centering
\includegraphics[width=8.5cm]{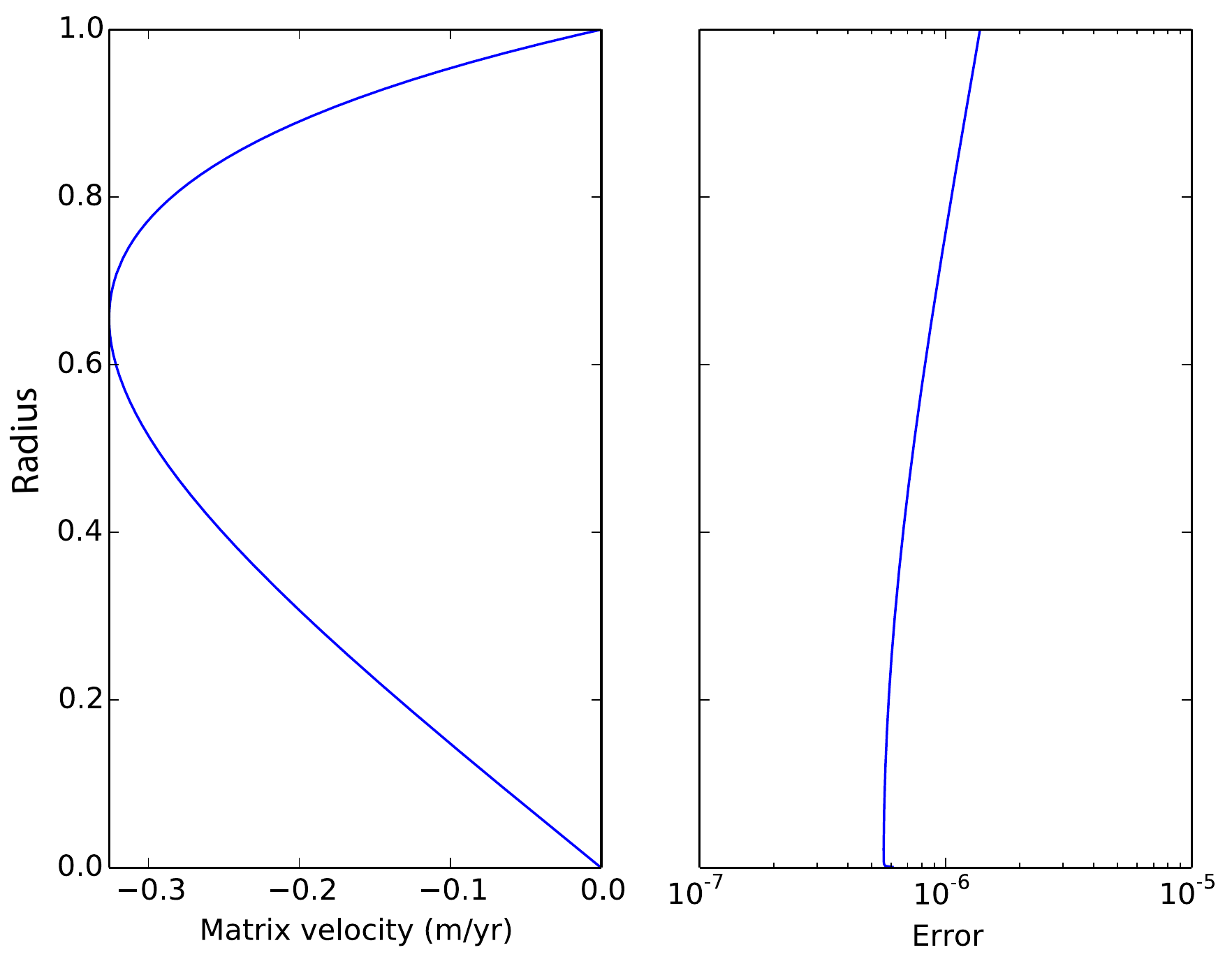}
\caption[Solid fluid separation: analytical versus numerical solutions.]{\textbf{a}: Analytical solution of the matrix velocity for a constant porosity profile (eqn \ref{eqn:ana_vsol}), with parameters: $R = 263$ km, $\phi_l=0.1$, $\overline{\rho}$ = 3260 kg/m$^3$, $\rho_f$ = 2900 $kg/m^3$, $L_c = 17.3$ km and a number of radial levels $nr=1000$. \textbf{b}: Relative error between the analytical solution and the numerical solution computed by a tridiagonal inversion and a backward Euler finite difference scheme.}
\label{fig:ana_num}
\end{figure}

The validity of the numerical resolution of equation (\ref{eqn:phi_adim}) was controlled by comparing it to an analytical solution. This is done in the case of a constant porosity profile, for which the matrix velocity equation (\ref{eqn:phi_adim}) reduces to:
\begin{equation}
 F \frac{\partial}{\partial r} \dfrac{1}{r^2} \frac{ \partial r^2 v_m}{\partial r} -\dfrac{R^2}{L_c^2} v_m = (1-\phi_0) r,
\label{eqn:vm_bis}
\end{equation}
whose analytical solution, considering an impermeable surface ($v_m(R)=0$), is:
\begin{align}
v_m= \dfrac{(1-\phi_0)L_c^2}{R^2}\left[\dfrac{(A-r) e^{\frac{r}{A}}- (A+r)e^{-\frac{r}{A}}}{B r^2} - r\right],
\label{eqn:ana_vsol}
\end{align}
where $A= L_c \sqrt{F}/R$ and $B = {(A-1) e^{1/A} - (A+1)e^{-1/A}}$.
Equation (\ref{eqn:vm_bis}) is solved numerically by a tridiagonal inversion method and backward Euler finite difference scheme. As shown in Figure\,\ref{fig:ana_num}, the numerical solution of the velocity equation is solved to a precision of $\sim$ 10$^{-6}$.

\subsection{Benchmark}\label{sec:Bench}
\begin{figure} [t]
\centering
\includegraphics[width=8.5cm]{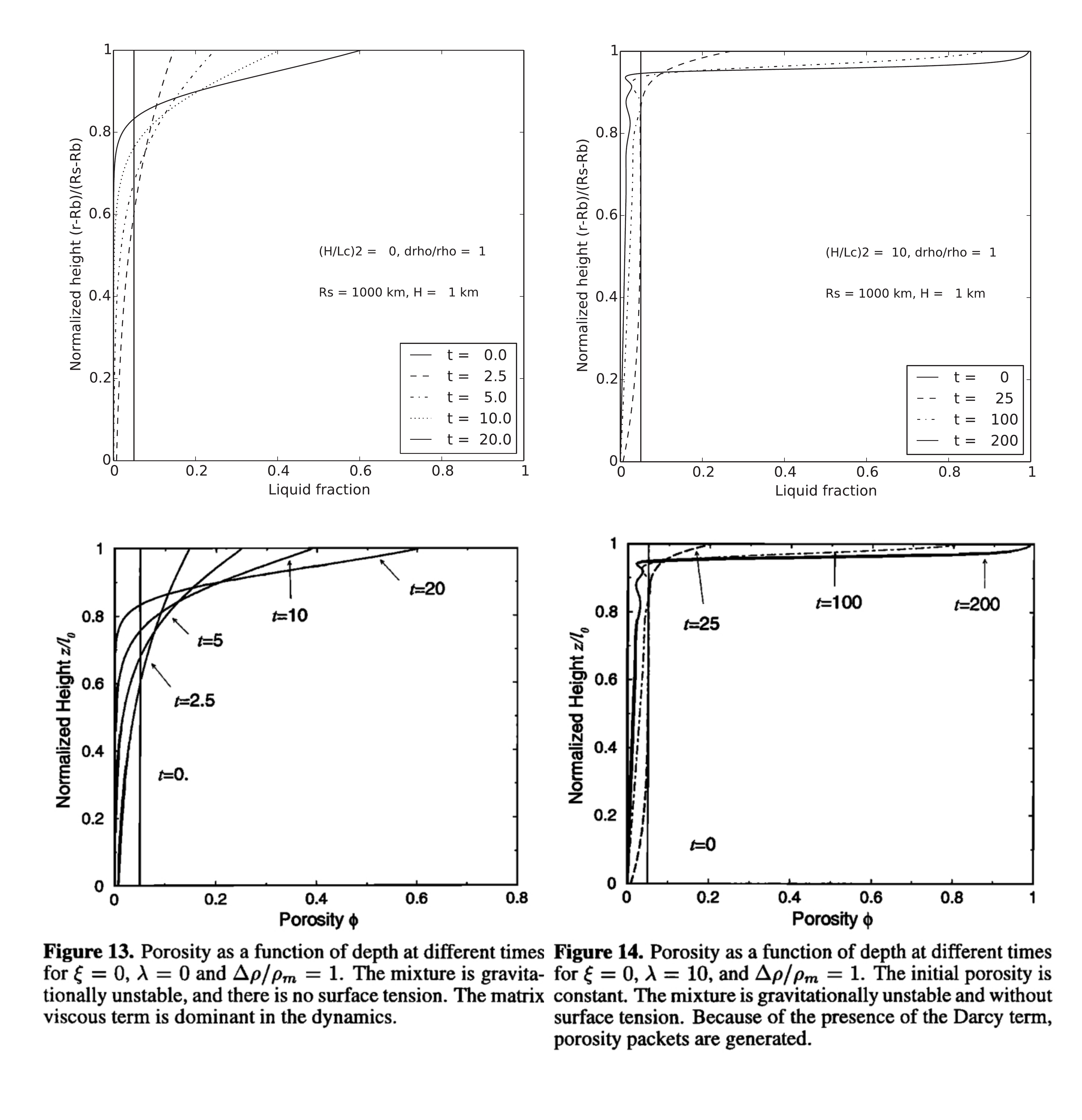}
\caption[Benchmark experiments]{(a) and (b) drainage experiments performed within an impermeable spherical shell of thickness $H = 1 km$, for an external radius of 1000 km. This aspect ratio is chosen in order to enable comparison with the cartesian solutions of \citep{Ricard2001} that are showed on the bottom row (Figure 13 \& 14). $(H/L_c)^2 \sim 0$ (a) and $(H/L_c)^2 = 10$ (b). The initial liquid fraction is 10\%. Our experiments are obtains using a tridiagonal inversion method and a backwad Euler finite different scheme for the velocity equation and a second order Runge-Kutta method and an upwind scheme for the advection equation. Time is normalized by $4\mu_m/(3\rho10H)$.}
\label{fig:benchmark_ricard}
\end{figure}
The numerical solution of liquid transport was benchmarked against two drainage experiments described in \citep{Ricard2001}. The separation between a low viscosity fluid and a highly viscous matrix, under the effect of gravity, is computed for an imposed constant initial liquid fraction. The authors used a slightly different formalism to which we adapted our code to allow the comparison of the experiments \citep[see][for more details]{Mizzon2015}.
Figure\,\ref{fig:benchmark_ricard} shows this comparison. The profiles of liquid fractions and the time evolution calculated here appear consistent with the numerical simulation of \cite{Ricard2001}. The dependency of the flow regime on the $R/L_c$ ratio is reproduced. When the size of the system is large compared to the compaction length, compaction waves are observed and when the size of the system is comparable to the compaction length, the liquid fraction is always a monotonically decreasing function of depth.

\bibliographystyle{elsarticle-harv} 
\bibliography{Biblio}

\end{document}